\documentclass[reqno,12pt]{article}
\usepackage{float}
\usepackage{ae} 
\usepackage[T1]{fontenc}
\usepackage[ansinew]{inputenc}
\usepackage{amsmath}
\usepackage{amssymb}
\usepackage{amsthm}
\newtheorem{theorem}{Theorem}
\usepackage{graphicx}
\usepackage[normalem]{ulem}
\usepackage{color}
\definecolor{darkblue}{cmyk}{0.9,0.9,0,0}
\usepackage[colorlinks=true,linkcolor=darkblue,citecolor=darkblue,urlcolor=darkblue]{hyperref}
\usepackage{epsfig}

 \newcommand{\badat}{\begin{alignedat}}
 \newcommand{\eadat}{\end{alignedat}}
 
 \def\be{\begin{equation}}
\def\ee{\end{equation}}

\usepackage[dvipsnames]{xcolor}
\usepackage{graphicx}

\newcommand{\beq}{\begin{equation}}
\newcommand{\eeq}{\end{equation}}
\newcommand\beqa{\begin{eqnarray}}
\newcommand\eeqa{\end{eqnarray}}
\newcommand\bea{\begin{array}}
\newcommand\eea{\end{array}}

\def\XXint#1#2#3{{\setbox0=\hbox{$#1{#2#3}{\int}$}
\vcenter{\hbox{$#2#3$}}\kern-.5\wd0}}

\newcommand{\nn}{\nonumber}

\newcommand{\neqa}{\nonumber\end{eqnarray}}
\newcommand{\la}[1]{\label{#1}}

\newcommand{\p}{\partial}

\renewcommand{\d}{\partial}

\newcommand{\re}{\relax{\rm I\kern-.18em R}}

\renewcommand{\sp}{p\hspace{-.40em}/}

\def\su2{{SU(2)}}

\def\spi{\relax{\rm \pi\kern-0.5em /}}
\def\sA{\relax{\rm A\kern-0.5em /}}
\def\sp{\relax{\rm p\kern-0.5em /}}
\def\sd{\relax{\rm \d\kern-0.5em /}}
\def\sk{\relax{\rm k\kern-0.5em /}}
\def\sn{\relax{\rm n\kern-0.5em /}}
\def\sl{\relax{\rm l\kern-0.5em /}}
\def\sP{\relax{\rm P\kern-0.7em /}}
\def\sBethe{\relax{\rm \Bethe\kern-0.5em /}}

\usepackage{varioref}
\usepackage{makeidx}
\makeindex

\newcommand\blfootnote[1]{%
  \begingroup
  \renewcommand\thefootnote{}\footnote{\hspace{-6mm}#1}%
  \addtocounter{footnote}{-1}%
  \endgroup
}

\usepackage{tikz}
\usetikzlibrary{positioning, arrows.meta, calc}
\usepackage{xcolor}

\newcommand{\secbox}[1]{%
    \fcolorbox{blue!70!black}{white}{%
        \textcolor{blue!80!black}{\footnotesize\bfseries \rule{0pt}{1.5ex}#1}%
    }%
}

\usepackage{amsmath}
\DeclareMathOperator{\csch}{csch}

\usepackage[english]{babel}
\begin{document}


\thispagestyle{empty}

\renewcommand{\thefootnote}{\fnsymbol{footnote}}
\setcounter{page}{1}
\setcounter{footnote}{0}
\setcounter{figure}{0}

\vspace{-0.4in}

\vspace{-0.4in}

\begin{center}
$$$$
{\Large\textbf{\mathversion{bold}
From Gross-Manes to Alday-Maldacena
}\par}
\vspace{1.0cm}

{ Leonardo Pipolo de Gioia,$^\text{\tiny 1}$ Sabrina Pasterski,$^\text{\tiny 2}$ Pedro Vieira$^\text{\tiny 1,\tiny 2}$}
\blfootnote{\tt{leonardo.de.gioia@gmail.com, pedrogvieira@gmail.com, spasterski@pitp.ca}}
\\ \vspace{1.2cm}
\footnotesize{
{\it $^\text{\tiny 1}$ICTP South American Institute for Fundamental Research, \\
IFT-UNESP, S\~ao Paulo, SP Brazil 01440-070}\\~\\
{\it $^\text{\tiny 2}$Perimeter Institute for Theoretical Physics,
Waterloo, Ontario N2L 2Y5, Canada  }\\
\vspace{4mm}
}
\end{center}

\begin{abstract}

Minimal surfaces govern the classical limit of several important observables in string theory. Two basic examples are the Gross--Manes saddle for high-energy string scattering in flat space and the Alday--Maldacena surface for gluon scattering at strong coupling in ${\cal N}=4$ SYM. In this paper we study a family of null polygonal minimal surfaces in AdS interpolating between these two regimes by moving the cusps from a small region deep in the bulk to the AdS boundary. We do this both numerically and -- in interesting simplifying limits -- analytically. In particular, we derive the leading AdS-radius correction around flat space and show that the resulting classical amplitude agrees with the recent prediction of Alday, Armanini, H\"aring, and Zhiboedov, supporting an underlying worldsheet description of their bootstrap construction.

\end{abstract}

\newpage

\setcounter{page}{1}
\renewcommand{\thefootnote}{\arabic{footnote}}
\setcounter{footnote}{0}

{
\tableofcontents
}

\newpage

\section{Introduction} 

\begin{figure}[t]
    \centering
\includegraphics[width=1\textwidth,angle=0,trim={0cm 0cm 0cm 0cm}, clip=true]{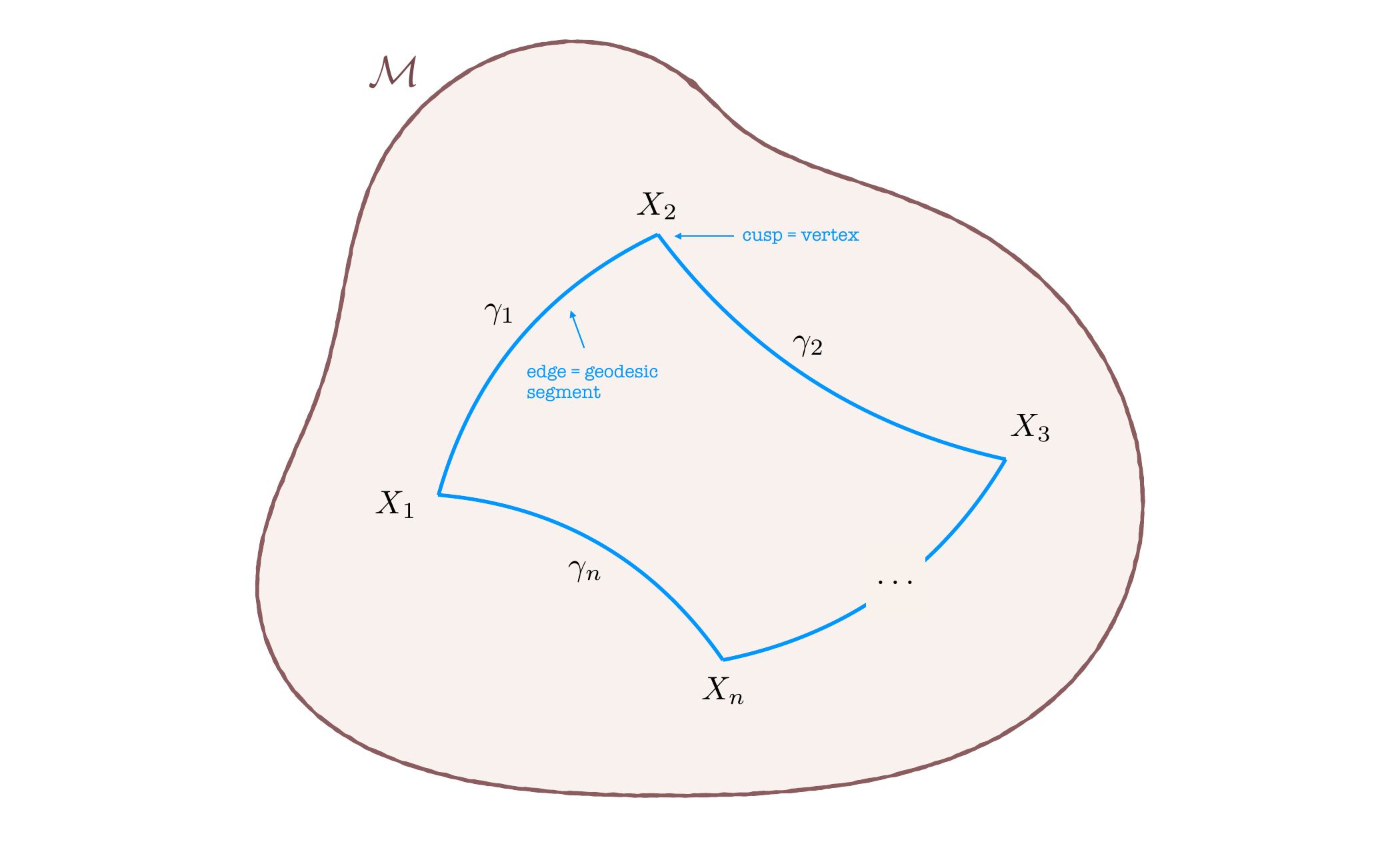}
    \caption{Polygon $=$ Sequence of geodesic segments $\gamma_i$ in some space $\mathcal{M}$.}
        \label{polygonCartoon}
\end{figure}

In this paper we embark on a purely geometric exercise inspired by string theory and holography: that of computing the areas of surfaces spanning null polygons in the
bulk of AdS.

We will use \textit{polygon} to denote a sequence of concatenated geodesic segments $\gamma_{i}$ -- with $i=1,\dots,n$ for an $n$-gon -- in some $d$-dimensional space (or space-time) $\mathcal{M}$. Each of the $n$ geodesics $\gamma_i$ connects two of the cusps $X_i$ and $X_{i+1}$ of the polygon, see figure \ref{polygonCartoon}. In particular, our polygons are not necessarily planar and refer only to the frame defined by these curves and not to the surface spanning it. Then we will use \textit{area of the polygon} to indicate the area of the two dimensional extremal surface bounded by the polygon. Sometimes this area might not be unique and in Lorentzian signature this extremal might not be a minima as we shall review soon.

Can we compute these areas? Despite the rather innocuous setup, it cannot be trivial. After all, any smooth curve can be approximated by a polygon with many tiny edges and finding the minimal surface for a general curve in a general space is of course a daunting task. {Indeed, finding an embedding $\mathbb{X} : \mathbb{D}\to \mathcal{M}$, where $\mathbb{D}$ is the unit disk in the complex plane, that extremizes the area functional and whose boundary $\mathbb{X}(S^1)\subset \mathcal{M}$ is a prescribed contour is a classical and well-studied problem in Riemannian geometry known as the Plateau problem \cite{plateau_survey}. In the mathematical literature, the Plateau problem concerns the existence and regularity of extremal surfaces, and in certain settings also their uniqueness, for given classes of boundary curves.

A seminal solution to the Plateau problem in $\mathbb{R}^3$ was given by Douglas \cite{rado1930plateau, Douglas1931}. His approach relies on working with conformal parametrizations of the worldsheet, which in physics corresponds to describing the string dynamics via the Polyakov action rather than the Nambu--Goto one. While this choice renders the bulk equations of motion linear --- reducing them to the Laplace equation --- it introduces the non-trivial Virasoro constraints, which restrict the allowed parametrizations of the boundary contour. Douglas showed that solving the Laplace equation with an arbitrary boundary parametrization, followed by a minimization over all such parametrizations, yields a solution to the Plateau problem.

Apart from proving existence, regularity, and uniqueness of extremal surfaces spanning certain classes of boundary curves, it is often possible to extract geometric information about the resulting surface. A particularly useful framework for this purpose is the Weierstrass parameterization of minimal surfaces \cite{weierstrass_rep}. While Douglas's method can be thought of as simplifying the imposition of boundary conditions through the use of Green's functions, at the price of rendering the Virasoro constraints non-trivial, the Weierstrass parameterization trivializes the Virasoro constraints, rendering the imposition of boundary conditions non-trivial. Nevertheless, the Weierstrass parameterization expresses the surface in terms of highly significant geometric data, most notably its Hopf differential $p(z)dz^2$, which encodes curvature information of the minimal surface.

A historically important application of the Weierstrass parameterization is the study of minimal surfaces spanning polygonal contours in $\mathbb{R}^3$. In a highly symmetric configuration known as the Schwarz quadrilateral, it is possible to derive an algebraic equation whose solution determines the surface's Hopf differential and, through the Weierstrass parameterization, yields the embedding explicitly \cite{schwarz_quad}. Similar equations can be solved numerically in much more general configurations, providing minimal conformal embeddings spanning polygonal contours in $\mathbb{R}^3$.} 

While the classical Plateau problem is firmly rooted in Riemannian geometry, our physical applications require studying embeddings in space-times with Lorentzian signature, such as $\mathbb{R}^{1,D-1}$ or AdS$_{d+1}$. This indefinite signature introduces important subtleties. In a Lorentzian manifold, extremizing the area functional does not strictly yield ``minimal'' surfaces: depending on the space-like or time-like nature of the variations, these extremal surfaces will act as saddle points or even maxima of the area functional \cite{cheng1976maximal,bartnik1982spacelike,umehara2006maximal}. Furthermore, boundary conditions involving null geodesic segments, such as the null polygons we consider here, are notoriously delicate \cite{akamine2021reflection}. Nevertheless, there are several advances in the mathematical literature regarding the Plateau problem in Lorentzian signature, see for example \cite{quien1985plateau,desideri2010plateau} and references therein. Despite these rigorous mathematical distinctions, it is standard practice in the high-energy physics literature to refer to these extremal surfaces loosely as ``minimal surfaces''. We will adopt this standard abuse of terminology throughout the paper.

Our medium term goal\footnote{Our short term goal -- in this paper -- will be to compute some areas \textit{by brute force}, by explicitly finding the minimal surfaces, in preparation to that more ambitious goal. } is to study particular space(-times) $\mathcal{M}$ for which the area problem is integrable and try to develop integrability based techniques for finding these areas \textit{without} finding the explicit form of the surface. A primary physical motivation for this problem is that, in certain regimes such as the high-energy limit of flat-space scattering, the string path integral is dominated by classical saddles corresponding to these extremal surfaces. Because the areas of these surfaces yield the on-shell action, their exponentiation directly computes important physical observables, ranging from scattering amplitudes in flat-space string theory to dual CFT observables in AdS/CFT.

Ultimately, one of the key generalizations of this minimal surface problem is to consider quantum vibrating strings for which the minimal areas represent only the leading classical result. The hope is that the integrability based results of these areas might exhibit interesting new mathematical structures which might hint at novel descriptions of string theory. In particular, we hope that these geometrical classical problems and their quantum extensions might help better understand some aspects of flat space string theory and flat space holography. Take $\mathcal{M}$ to be Anti de Sitter space for example. By playing with the shape of the polygon we can nicely interpolate between boundary quantities when taking the cusps $X_i$ to approach the boundary of $AdS$ and flat space areas by taking the cusps $X_i$ to approach each other in a small bulk region inside $AdS$. This fact that these minimal polygon surfaces are not anchored at the boundary is of course the main novelty here; minimal surfaces ending at the boundary of $AdS$ have of course been studied in great  detail in the past, see e.g. \cite{Maldacena:1998im, Gross:1998gk, Drukker:1999zq,Kruczenski:2002fb,Alday:2007hr} and \cite{Alday:2009yn, Alday:2010ub, Alday:2010ku, alday2011thermodynamic}.

The first group of references are famous examples of area computations based on the explicit computation of a minimal surface; the second group of references are integrability based computations of minimal areas for which one could not find the minimal surface but for which it was nonetheless possible to find the minimal area. 

\begin{figure}[t]
    \centering
\includegraphics[width=\textwidth]{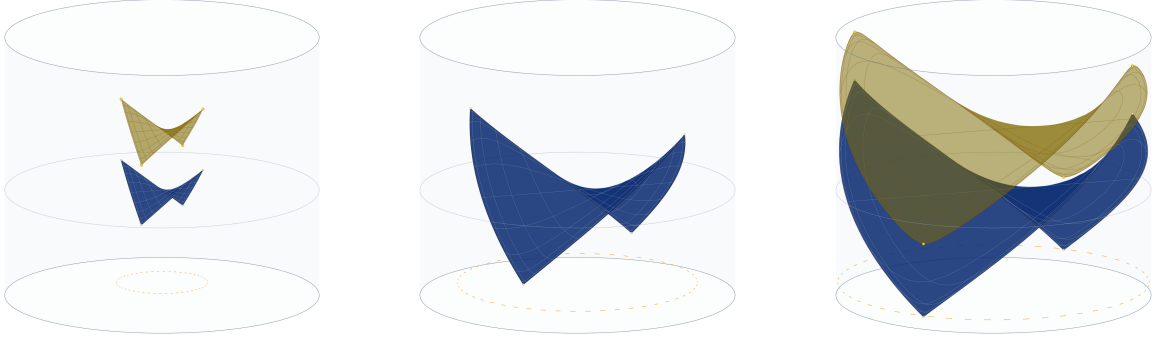}
    \caption{{\textbf{Left:}} Flat Space Limit numerical surface for $S=T\ll 1$. We match with the high-energy surfaces of Gross and Manes. {\textbf{Middle:}} Intermediate configuration of our numerical surface for $S = T$. {\textbf{Right:}} Large Polygon Limit $S=T\gg 1$ numerical surface. In this regime we match with the Alday-Maldacena null quadrilateral. Our numerical surfaces are shown in blue and the predicted Gross-Manes and Alday-Maldacena surfaces are shown in yellow and displaced in $\tau$ for better comparison. 
    }
        \label{fig:interpolation}
\end{figure}

We will study here $n$-gons with null edges in $\mathcal{M}= AdS_d$. If the cusps approach the boundary of $AdS$ these become null polygons at the boundary and -- in an AdS/CFT setup -- we would be computing null polygonal Wilson loops at strong coupling. These were related to gluon scattering amplitudes in some gauge theories by Alday and Maldacena \cite{Alday:2009yn}. If, on the contrary, all the cusps of the polygon approach each other, in some region much smaller than the AdS curvature $L$, we will recover the flat space result for $\mathcal{M}=\mathbb{R}^{1,d-1}$. The flat space minimal area problem was beautifully studied by McGreevy and Sever \cite{McGreevy:2007kt} who -- following Alday and Maldacena -- pointed out that the minimal surfaces found by Gross and Manes \cite{Gross:1989ge} (see also \cite{Gross:1987kza}) in the context of high energy string scattering become precisely these null polygonal minimal surfaces under a generalized $T$-duality transformation. (We review this story in section \ref{sec:flatlimit}.) The interpolation between these two regimes is depicted in figure \ref{fig:interpolation}.
A closely related related important motivation for studying these minimal areas is scattering in the Coulomb branch of $\mathcal{N}=4$ SYM \cite{McGreevy:2008zy,Alday:2009zm}. These polygonal minimal areas compute the strong coupling classical limit of these objects. A true showpiece is the recent paper \cite{Alday:2025pmg} by Alday, Armanini, H\"aring and Zhiboedov where scattering in the Coulomb branch of $\mathcal{N}=4$ SYM was studied at weak, strong and even finite coupling combining integrability and bootstrap techniques. 

\begin{figure}[t]
    \centering
\includegraphics[width=1\textwidth,angle=0,trim={0cm 0cm 0cm 0cm}, clip=true]{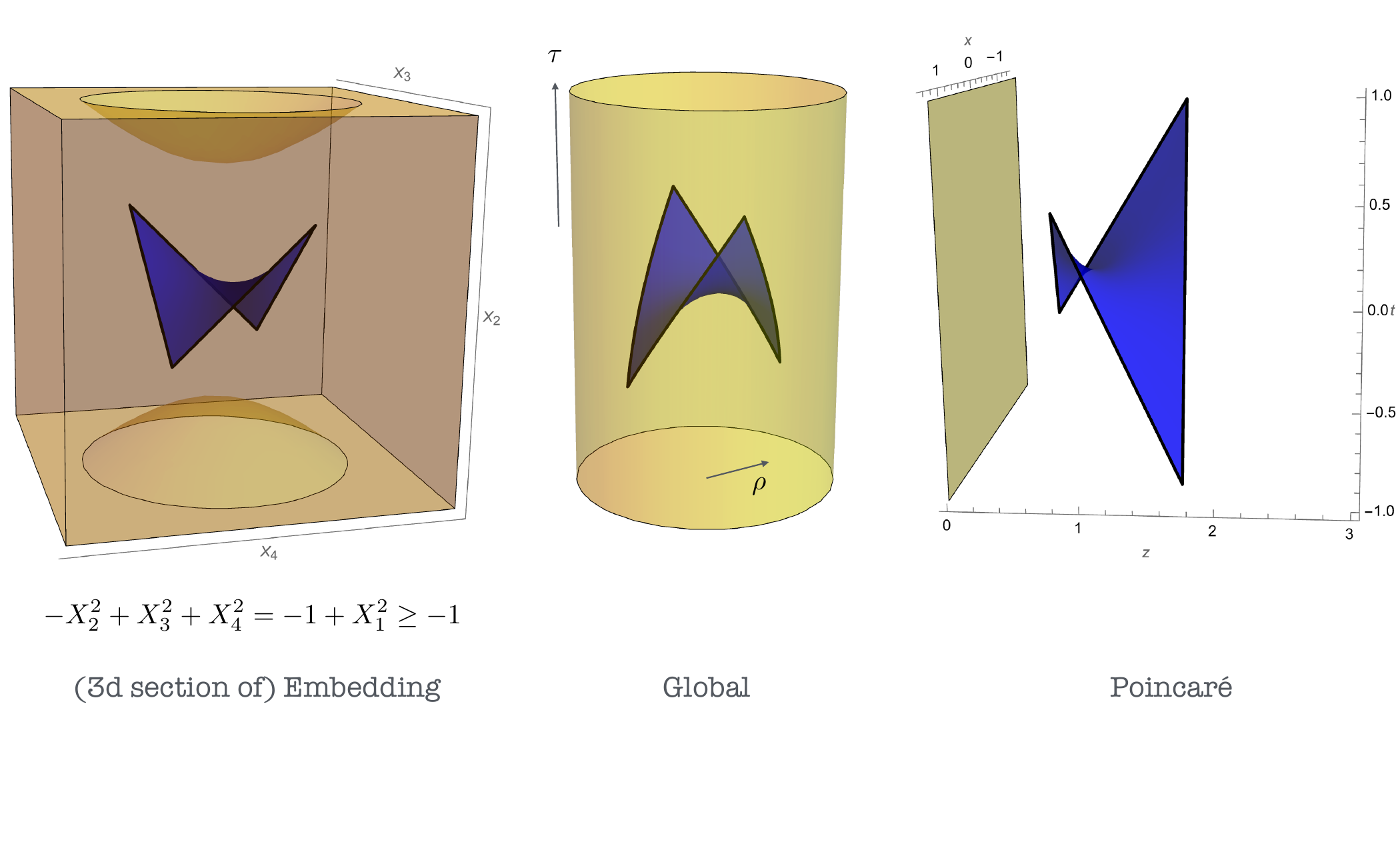}
\vspace{-2cm}
    \caption{\textbf{Left}: Minimal surface in AdS represented as a hyperboloid in higher-dimensional space. \textbf{Middle}: Minimal surface in AdS represented in global coordinates. \textbf{Right}: Minimal surface in AdS in the Poincar\'e patch. Note that for the Poincar\'e patch representation in AdS$_3$, a non-trivial surface cannot be at fixed $z$, but can be if it is instead embedded in AdS$_4$.}
        \label{DifCoordinates}
\end{figure}

The simplest area in this setup is that of the $4-gon$ or \textit{null rectangle}. This polygon is given by four cusps $X_i$ connected by null geodesics. AdS isometries relate different squares preserving the distances $(X_i-X_j)^2$ between them. Since all edges are null, the area is therefore a function of the diagonal lengths only, a function of two variables 
\beq
\texttt{Area}(S,T) = L^2 A(S,T)
\eeq
with the dimensionless variables $S=(X_1-X_3)^2/L^2$ and $T=(X_2-X_4)^2/L^2$. Here we use embedding coordinates $X_i$ with $X_i^2=-L^2$. Of course, we visualize the same polygon in embedding, global, Poincare or any other coordinates, see figure \ref{DifCoordinates}. In this paper we will study $A(S,T)$ analytically in several limits by studying the minimal polygonal surface in several limits such as $S,T\ll 1$ (flat space GM limit); $S,T \gg 1$ (AdS AM limit); and $S\gg T$ (Regge limit). We will also study it numerically in the special kinematics $S=T$ with the parameter interpolating from small to large values. As stated above, one of the goals of these explorations is to collect data/predictions for a future integrability based analysis. Complementarily, we also discuss some simple novel results about null $n$-gons ($n>4$) in flat space which we hope to explore further in the future. 

In the flat-space limit, one recovers the Gross--Manes minimal surface \cite{Gross:1989ge}, which is bounded by a null rectangle deep in the bulk of AdS. We find the area in this Gross--Manes limit to be
\begin{equation}
    \texttt{Area}_{GM} = \frac{1}{2\pi}\left[ (S+T)\log(S+T) - S\log S - T\log T\right].
\end{equation}
We can study the leading correction to the Gross-Manes area in two complementary limits. First, in the Regge regime $S\to\infty$, followed by the small-$T$ expansion, the coefficient of $\log S$ receives the correction
\begin{equation}\texttt{Area}(S,T)\simeq\left(\frac{T}{2\pi}+\alpha T^2+\cdots\right)\log S .
\end{equation}Second, when both invariants are small, the area in the symmetric kinematics $S=T$ receives a correction
\begin{equation}
    \texttt{Area}(T,T)\simeq\frac{T}{2\pi}\log 4+\gamma T^2+\cdots
\end{equation}
Thus, while $\alpha$ describes the leading curvature correction in the Regge kinematics, $\gamma$ measures the first departure from the flat-space Gross-Manes area along the symmetric trajectory $S=T$.

On the other hand, for large null rectangles ending on the AdS boundary, one finds the Alday--Maldacena minimal surface \cite{Alday:2009yn}, whose area plus corrections -- in our geometric regularization -- is given by 
\begin{equation}
    \texttt{Area}_{AM}(S,T) =\frac{1}{2} \log(S)\log(T) + \beta(\log(S)+\log(T)) + C 
\end{equation}
where $\beta$ has been computed from the Regge limit and we did not manage to compute $C$ analytically.\footnote{Numerically, we find $C \simeq 1.45$.} The leading term, $\frac{1}{2} \log(S)\log(T)$ agrees with the leading term in Alday-Maldacena as it ought to. The single logs and the constant are scheme dependent and do not need to match.  

\begin{figure}[t]
    \centering
    \begin{tikzpicture}[
        >=Stealth,
        box/.style={
            align=center,
            font=\footnotesize,
            inner sep=4pt,
            text width=5.2cm
        },
        topbox/.style={
            box,
            text width=7.4cm
        },
        arrow/.style={
            ->,
            thick,
            color=black!75
        },
        bidir/.style={
            <->,
            thick,
            color=black!75
        },
        lbl/.style={
            font=\scriptsize,
            align=center,
            inner sep=2pt,
            fill=white,
            text=black!85
        }
    ]


    \node[topbox] (sec2) at (0,0) {
        \textbf{Regge Limit Interpolation} \\
        $S\to\infty$ at fixed $T$ \\
        Small $T$: predicts $\alpha$;
        large $T$: predicts $\beta$ \\[.8ex]
        \secbox{\hyperref[sec:regge]{Section 2}}
    };


    \node[box] (sec3) at (-4.7,-4.4) {
        \textbf{Nambu--Goto Numerics} \\
        Symmetric kinematics $S=T$ \\
        Large invariants: tests $\beta$ \\
        Small invariants: determines $\gamma$ \\[.8ex]
        \secbox{
            \hyperref[sec:nambu-goto-numerics]{Section 3}
        }
    };


    \node[box] (sec4) at (4.7,-4.4) {
        \textbf{Flat-Space Curvature Correction} \\
        Conformal gauge, general $(S,T)$ \\
        Reproduces $\alpha$ in the Regge limit \\
        and $\gamma$ for $S=T$ \\[.8ex]
        \secbox{
            \hyperref[sec:flat-space-corrections]{Section 4}
        }
    };


    \node[box] (amp) at (4.7,-8.3) {
        \textbf{AdS Amplitude Comparison} \\
        Classical high-energy limit \\
        Match for the full correction
        $\delta\texttt{Area}(S,T)$ \\[.8ex]
        \secbox{
            \hyperref[app:correction-match]{Appendix D}
        }
    };


    \draw[arrow]
        (sec2.south west)
        to[out=210,in=90]
        node[lbl,left=3mm,pos=0.48] {
            Prediction for $\beta$ \\
            tested numerically
        }
        (sec3.north);

    \draw[arrow]
        (sec2.south east)
        to[out=330,in=90]
        node[lbl,right=3mm,pos=0.48] {
            Prediction for $\alpha$ \\
            reproduced analytically
        }
        (sec4.north);

    \draw[bidir]
        (sec3.east) --
        node[lbl,above=1mm] {
            Independent agreement \\
            on $\gamma$
        }
        (sec4.west);

    \draw[arrow]
        (sec4.south) --
        node[lbl,right=1mm] {
            Agreement with \\
            AdS amplitudes
        }
        (amp.north);

    \end{tikzpicture}

    \caption{
        Roadmap of the complementary computations and checks.
        The Regge solution predicts $\alpha$ and $\beta$;
        the Nambu--Goto numerics tests $\beta$ and determines
        $\gamma$; and the conformal-gauge expansion reproduces
        $\alpha$ and $\gamma$ and matches the AdS-amplitude
        correction.
    }
    \label{fig:roadmap}
\end{figure}
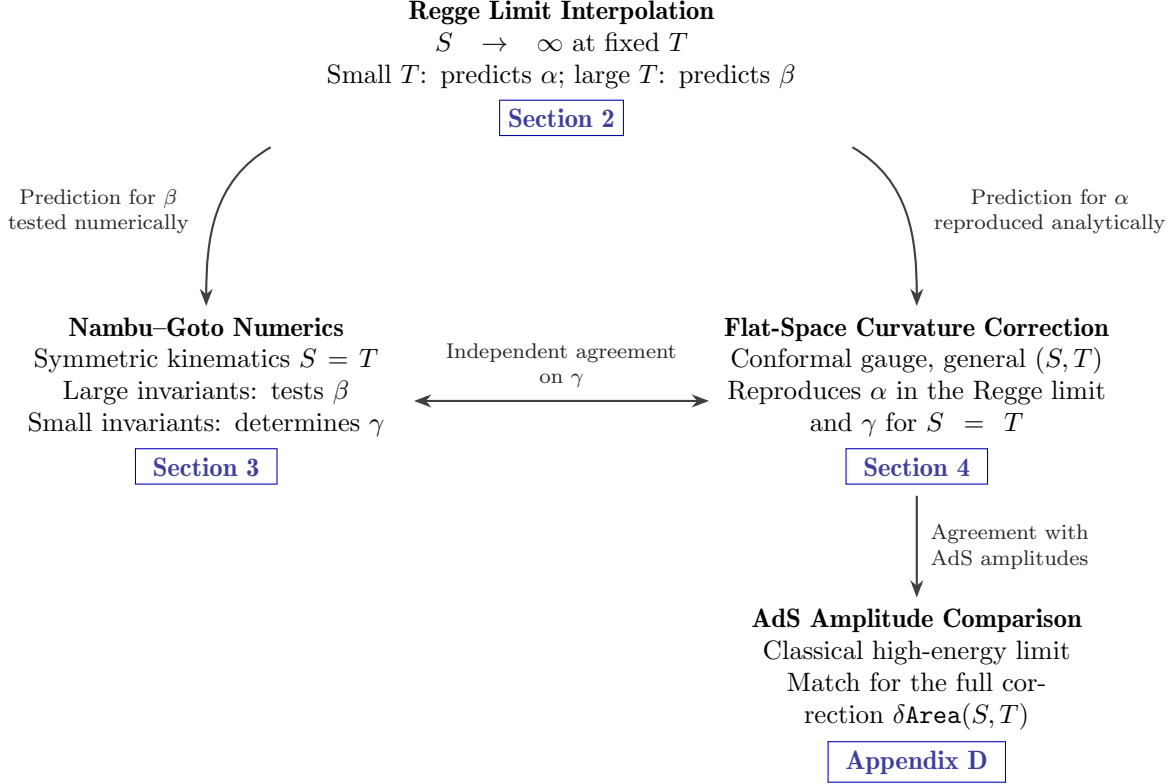

This paper is organized as follows, see figure \ref{fig:roadmap}. In Section \ref{sec:regge}, we show that in the Regge limit and at large $S$, the minimal area takes the form
\begin{equation}
    \texttt{Area}_{\text{Regge}} = f(T)\log S ,
\end{equation}
where the function $f(T)$ has the asymptotic behaviors
\begin{equation}
    f(T) = \begin{cases}
    \tfrac{T}{2\pi}, & T \to 0,\\
    \tfrac{\log T}{2}, & T \to \infty.
    \end{cases}
\end{equation}
The small-$T$ regime, therefore, reproduces the leading large-$S$ behavior of the Gross--Manes flat-space solution, while the large-$T$ regime reproduces the corresponding single-log contribution in the Alday--Maldacena regime. Moreover, the asymptotic expansion of $f(T)$ predicts both $\alpha$ and $\beta$, namely as $T\to 0$
\begin{equation}
    f(T) \simeq \frac{T}{2\pi} + \alpha T^2, \qquad \alpha = -\dfrac{15+2\pi^2}{48\pi^3},\qquad T \to 0,
\end{equation}
while as $T\to \infty$
\begin{equation}
    f(T) \simeq \frac{\log T}{2} + \beta,\qquad \beta = \log(1+\sqrt{2})-\sqrt{2}, \qquad T \to \infty,
\end{equation}
We then turn to the numerical Nambu--Goto analysis in
Section~\ref{sec:nambu-goto-numerics}, which probes both ends of the
interpolation. Near the Alday--Maldacena regime, the numerics test the
Regge-limit prediction for $\beta$, while near the symmetric
Gross--Manes regime, they provide a numerical determination of $\gamma$.
In Section~\ref{sec:flat-space-corrections}, we independently derive the leading curvature correction around flat space in conformal gauge for
general kinematics. Its symmetric and Regge limits reproduce $\gamma$
and $\alpha$, respectively, yielding a nontrivial agreement among the
numerical, Regge-limit, and conformal-gauge analyses. We further show
that this correction to the area is compatible with the classical
high-energy limit of the AdS-radius correction to the Veneziano amplitude
recently discussed in \cite{Alday:2025pmg}.

\section{
Regge Limit Interpolation}\label{sec:regge}

\subsection{Coordinate Conventions}

We want to study minimal surfaces ending on null polygons in Lorentzian AdS$_3$, viewed as the universal covering space of the hyperboloid in $\mathbb{R}^{2,2}$ defined by
\begin{equation}
    X^2 = -L^2,
\end{equation}
where we assume the signature convention $\eta = \operatorname{diag}(-1,-1,1,1)$ for the metric of embedding space. We use embedding space coordinates $X = (X^0,X^1, X^2,X^3)$ in what follows. Focusing on the case of a null quadrilateral, the polygon has four edges $X_i$ and, as described in the introduction, is described by two parameters
\begin{equation}
    S = \frac{(X_1-X_3)^2}{L^2},\quad T = \frac{(X_2-X_4)^2}{L^2}.
\end{equation}
These are dimensionless numbers so that small/big $(S,T)$ are small/big polygons (as compared to the AdS scale). 

\begin{figure}[h!]\label{fig:regge-fig}
        \centering
\includegraphics[width=.8\linewidth]{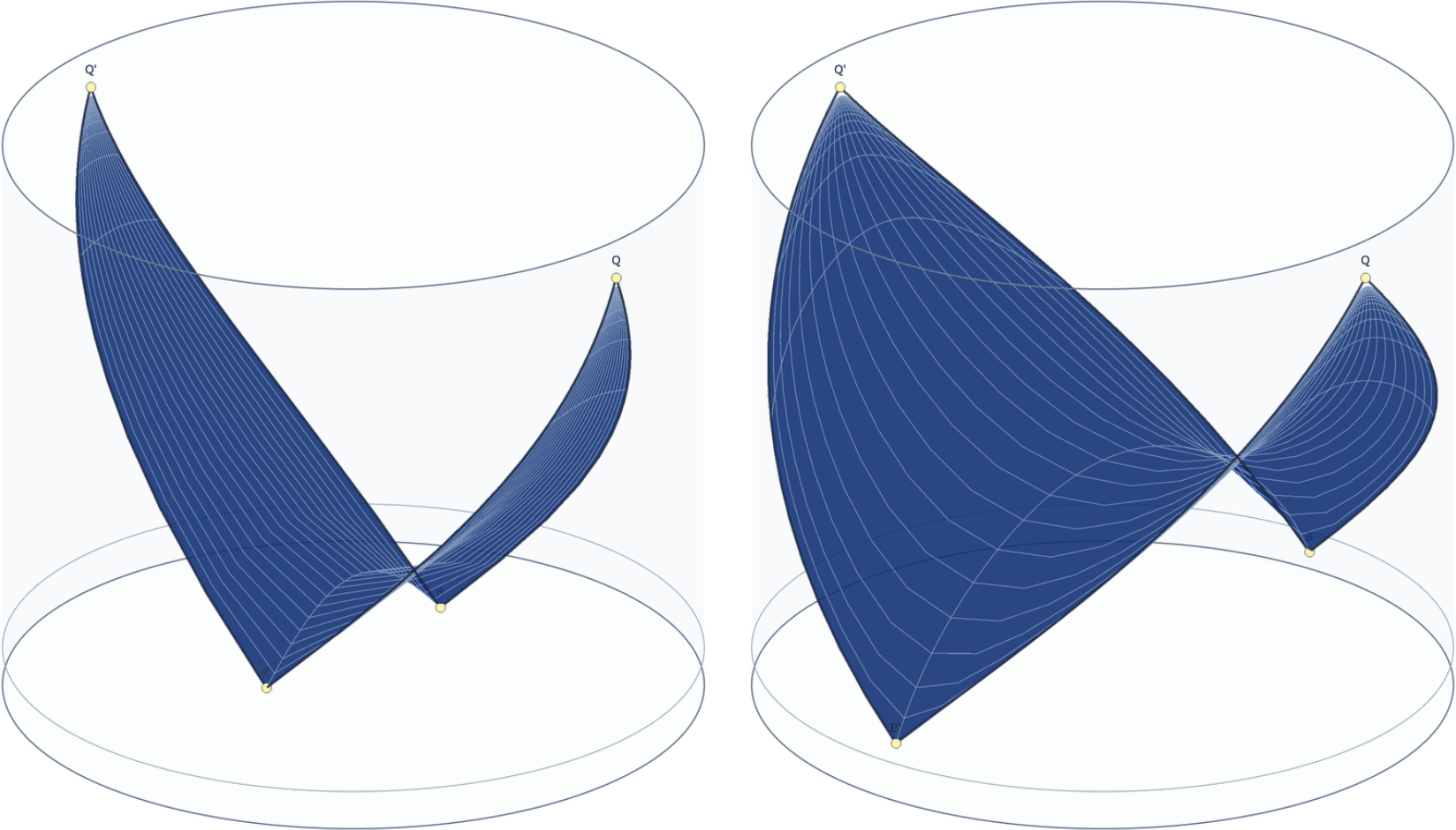}
        \caption{The minimal surface obtained in this section in the Regge limit in two different regimes. On the left, for small $T$, we have two of the points on the bulk, and on the right, for large $T$, we see that these two points go to the boundary.}
        \label{fig:placeholder}
    \end{figure}

In this section we are interested in the Regge limit configuration $S\to \infty$ with $T$ fixed. In this configuration, two of the corners of the polygon, say $Q$ and $Q'$, will be located at the boundary of AdS$_3$, while the other two corners will be inside the bulk. From each of the corners on the boundary, two null geodesics emanate, connecting to the two bulk points to form the polygon. This configuration is depicted in figure \ref{fig:regge-fig}. 
To deal with this configuration, it is useful to select a coordinate system that is adapted to the lightsheet of the boundary points. Given a boundary point $Q$, its lightsheet is defined by
\begin{equation}
    X^2 = -L^2,\quad X\cdot Q =0.
\end{equation}
For definiteness, we are going to take the two boundary points to be $Q = (0,1,1,0)$ and $Q'= (0,1,-1,0)$. The lightsheet of $Q$ can be parameterized by
\begin{equation}
    X^\mu (\lambda,\eta) = L (\cosh \eta,\lambda,\lambda,\sinh\eta).
\end{equation}
We notice that as $\lambda\to \infty$, factoring $\lambda$ out, we can recognize that all points $X^\mu(\lambda,\eta)$ go to the same boundary point $Q$ irrespective of the value of $\eta$, which becomes a cusp of the lightsheet. The lightsheet of $Q'$ is parameterized by the same formula upon flipping the sign of the $X^2$ component. The frame is defined by adjoining to $Q$ and $Q'$ two bulk points $B$ and $B'$ ensuring that all four vertices are connected by null geodesics. In order to describe a minimal embedding ending in such a frame, we select a coordinate system for AdS$_3$ that interpolates between the two lightsheets. One such set of coordinates is given by
\begin{equation}\label{eq:ads-embedding-regge}
    X^\mu = \dfrac{L}{\sqrt{1+2\lambda z + z^2}}(\cosh\eta,\lambda+z,\lambda,\sinh\eta)
\end{equation}
We then study a family of surfaces described by
\begin{equation}\label{eq:surface-regge-ansatz}
    z(\lambda,\eta) = -\lambda + \sqrt{\lambda^2 + f(\eta)},
\end{equation}
where we take $\eta \in (-R,R)$ and $\lambda\in\mathbb R$ and regulate the two asymptotic ends symmetrically by restricting $\lambda\in(-\Lambda,\Lambda)$. The principal branch in \eqref{eq:surface-regge-ansatz} automatically interpolates between the two lightsheets. Indeed, when $f(\eta)=0$, one has $z=-\lambda+|\lambda|$. For $\lambda>0$, the embedding reduces to
\begin{equation}
X^\mu=L(\cosh\eta,\lambda,\lambda,\sinh\eta),
\end{equation}
which lies on the lightsheet of $Q$. For $\lambda<0$, writing $\rho=-\lambda>0$, it becomes
\begin{equation}
X^\mu=L(\cosh\eta,\rho,-\rho,\sinh\eta),
\end{equation}
which lies on the lightsheet of $Q'$. At $\eta=\pm R$, these two null rays meet at the bulk points $B_\pm=L(\cosh R,0,0,\pm\sinh R)$, so that $T=4\sinh^2R$. On the other hand, $\Lambda$ regulates the two boundary cusps and controls the Regge limit and we can identify that $S=4\Lambda^2$ so that the Regge limit $S\to \infty$ corresponds to the $\Lambda\to \infty$ limit in which the cutoff is removed.

Constructing the Nambu-Goto lagrangian for this ansatz, we find that dependence in $\lambda$ is trivial enough so that the action depends on the chosen cutoff $\Lambda$ simply through a prefactor, giving rise to an effective lagrangian that governs the function $f(\eta)$
\begin{equation}
    \texttt{Area}_{\rm Regge} =2\operatorname{arcsinh}(\Lambda) \int d\eta\  \mathbb{L},\quad \mathbb{L} = \frac{ \sqrt{4 f(\eta ) (f(\eta )+1)-f'(\eta )^2}}{2 (f(\eta )+1)^{3/2}}.
\end{equation}
In terms of the $S$ invariant, in the $S\to \infty$ limit, this is
\begin{equation}
    \texttt{Area}_{\rm Regge} \simeq \log S \int d\eta\  \mathbb{L},
\end{equation}
and so we see that in this regime we are only capturing the leading log at large $S$, but with the complete dependence in $T$. As $T\to 0$, its leading-log expansion will be shown later to agree with the Gross-Manes result and its first AdS correction, whereas as $T\to\infty$ the polygon approaches the AdS boundary and the result matches the Alday-Maldacena regime. In what follows, we are going to discuss the solution to the ODE that governs $f(\eta)$ and the $T$ dependence of the area that it implies.

\subsection{A Regge Solution}

The Lagrangian $\mathbb{L}$ admits a conserved quantity that allows for the determination of $f(\eta)$ once initial conditions $f(0)$ and $f'(0)$ are prescribed. Relegating the derivation to Appendix \ref{app:regge-derivation}, the inverse function $\eta(f)$ with boundary conditions $f(0)=f_0$ and $f'(0)=0$ is given by
\begin{equation}
    \eta(f) = -\dfrac{\Pi\left(\frac{1}{1+f_0},\arcsin\sqrt{\frac{f(1+f_0)}{1+f}}\ \bigg|\ \frac{1}{f_0}\right)-\Pi\left(\frac{1}{1+f_0},\arcsin \sqrt{f_0}\ \bigg|\ \frac{1}{f_0}\right)}{\sqrt{1+f_0}}
\end{equation}
where $\Pi(n,\phi \ |\ m)$ is the incomplete elliptic integral
\begin{equation}
    \Pi(n,\phi \ |\ m) = \int_0^\phi \frac{d\theta}{(1-n\sin^2\theta)\sqrt{1-m\sin^2\theta}} \,.
\end{equation}
The data of the frame $S$ and $T$, or equivalently, $R$ and $\Lambda$, can now be extracted as follows: $\Lambda$ is defined to be the cutoff on the parameter $\lambda$, so that $S$ is truly a choice. On the other hand, $R$ is determined so that $f(R)=0$, or equivalently, $R =\eta(0)$. This determines $T$ to be
\begin{equation}
    T = 4\sinh^2 R,\quad R \equiv 
        R = \frac{\sqrt{f_0}}{\sqrt{1+f_0}}\Pi\left(\frac{f_0}{1+f_0}\ \bigg|\ f_0\right).    
\end{equation}
We plot $T$ as a function of $f_0$ on Figure \ref{fig:regge-T}.
\begin{figure}[t]\label{fig:regge-T}
    \centering
    \includegraphics[width=0.9\linewidth]{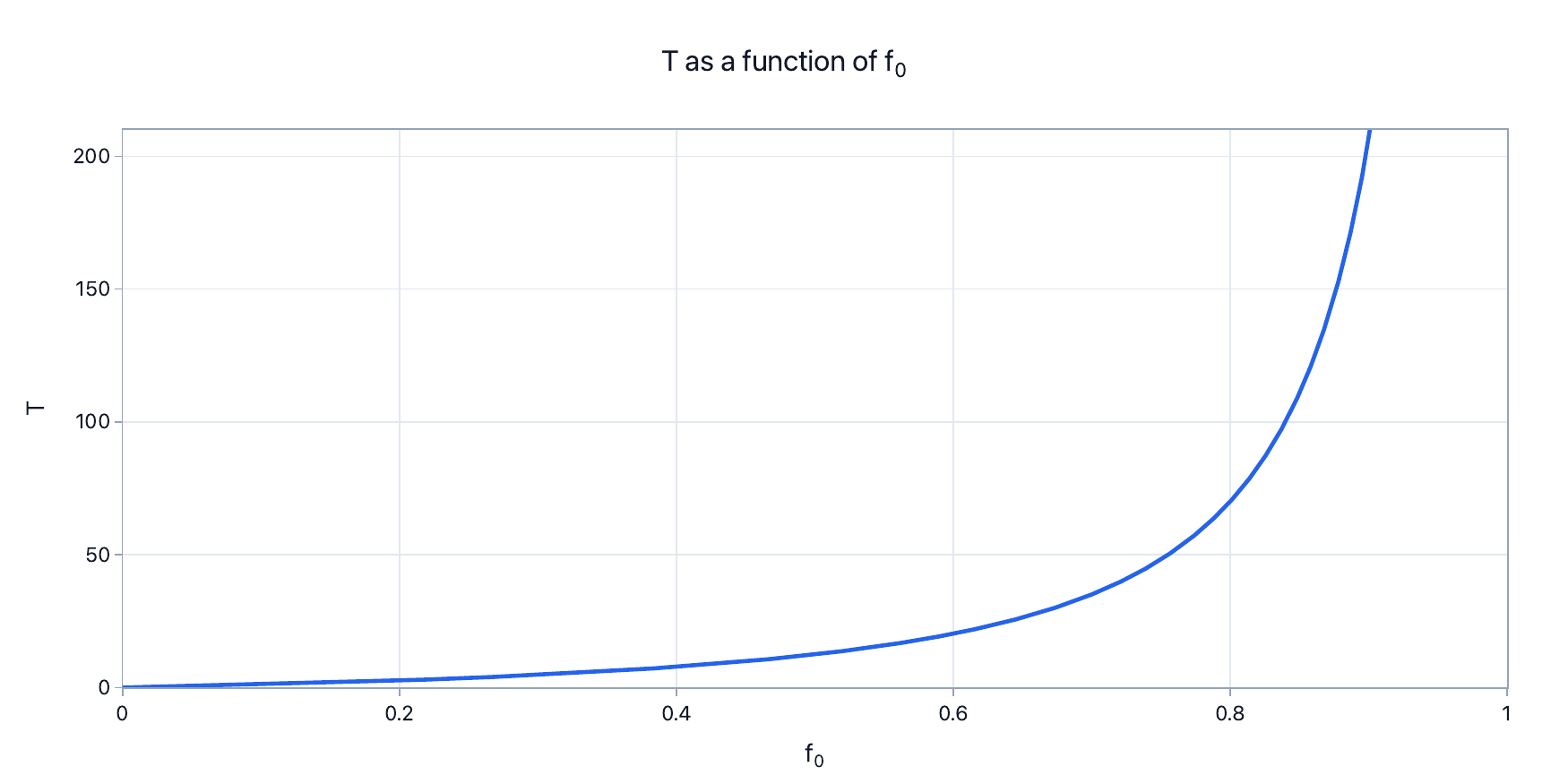}
    \caption{$T$ as a function of $f_0$}
    \label{fig:regge-T}
\end{figure}
The two interesting regimes of interpolation, namely $T\to 0$ and $T\to \infty$ can be reproduced by taking $f_0\to 0$ and $f_0\to 1$.

\begin{figure}[t]
    \centering
    \includegraphics[width=\linewidth]{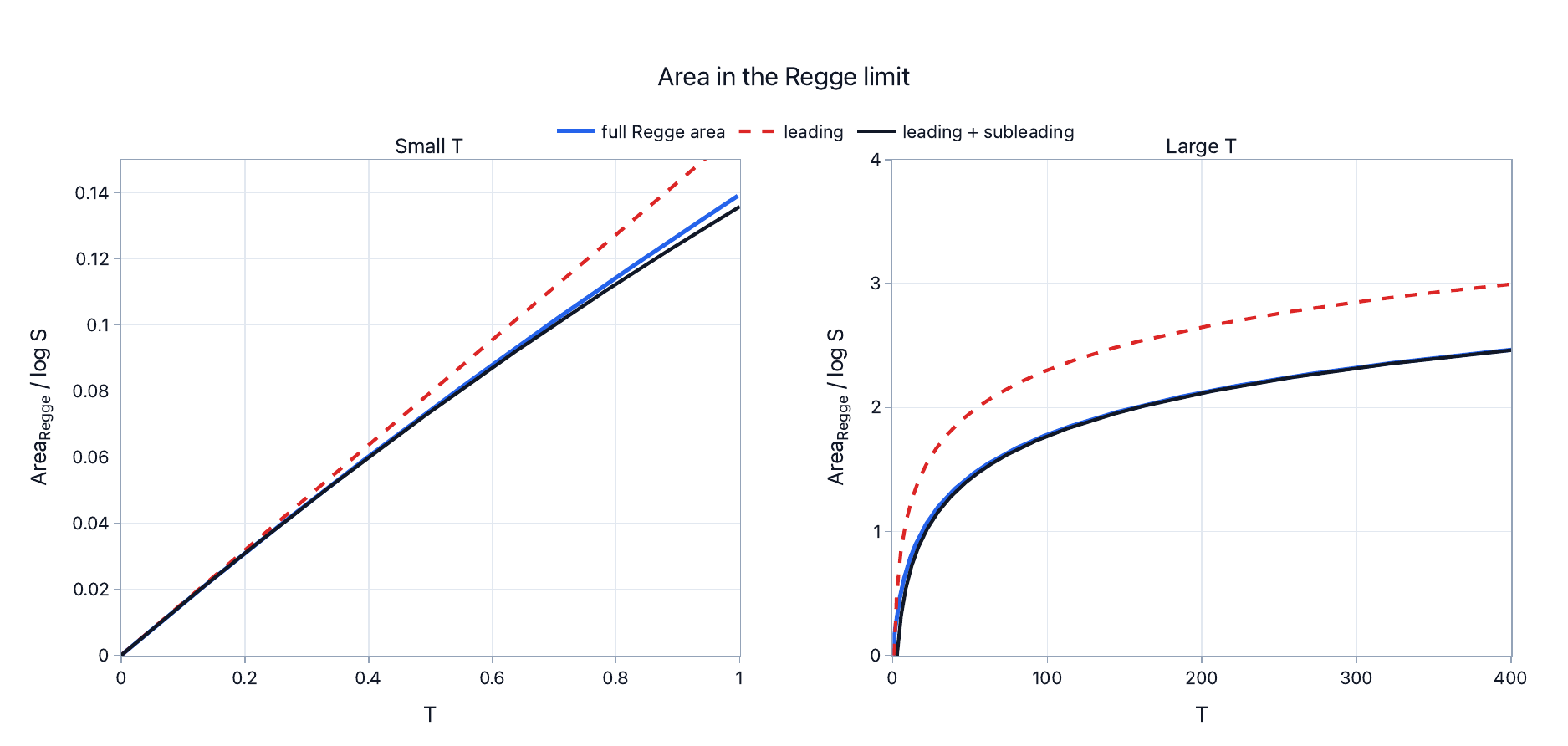}
    \caption{$\texttt{Area}$ in the Regge limit. \textbf{Left}: the area at small $T$. \textbf{Right}: the area at large $T$. In both plots, the blue curve is the full area result in the Regge limit \eqref{eq:area-regge}, the red curve is the leading term and the black curve the sum of the leading and subleading terms, in the corresponding regimes.}
    \label{fig:regge-area-expansions}
\end{figure}

The area can then be found by integrating the Lagrangian over $f$ instead of $\eta$ by a mere change of variables and eliminating $f'(\eta)$ in favor of $f(\eta)$ and $f_0$ using energy conservation. The resulting area can be computed and simplified as a function of $f_0$ into 
\begin{equation}\label{eq:area-regge}
    \texttt{Area}_{\text{Regge}} = \dfrac{2(K(f_0)-E(f_0))}{\sqrt{1+f_0}}\times \log S
\end{equation}
If we expand the area to subleading order in small $T$ we obtain the result
\begin{equation}\label{ReggeGM}
    \texttt{Area}_{\rm Regge}\approx \left(\frac{T}{2\pi} + \alpha T^2\right)\times \log S,\quad \alpha = -\frac{15+2\pi^2}{48\pi^3},\quad T\ll 1.
\end{equation}
The same leading log expansion in $S$ up to second order in $T$, with the same coefficient $\alpha$, appears when we study the AdS corrected area of the Gross-Manes surface in the Regge limit, as we shall show in section \ref{sec:flat-space-corrections}. On the other hand, at large $T$, and hence in an expansion around $f_0\approx 1$, the full bounding polygon asymptotes to the AdS boundary and the result should be compared to the Alday-Maldacena expectation. We find that the area has the behavior
\begin{equation}
    \texttt{Area}_{\rm Regge} \approx \left(\frac{\log T}{2}+\beta\right)\times \log S,\quad \beta = \log(1+\sqrt{2})-\sqrt{2},\quad T\gg 1.
\end{equation}
In figure \ref{fig:regge-area-expansions} we plot the full area compared to the leading and leading plus subleading terms both at small and large $T$, with the given values for $\alpha$ and $\beta$.

\section{A Nambu-Goto Story}\label{sec:nambu-goto-numerics}

\subsection{Coordinate Conventions}

In this section we will construct a numerical solution to the Nambu-Goto equations that will allow us to extract and compare both predictions of the Regge limit. To that end, however, it is useful to switch to a different coordinate system than the one we used in the previous section. We start by parameterizing AdS$_3$ in global coordinates
\begin{equation}\label{eq:Xemb}
X^\mu=L(\cosh r\cos\tau,\cosh r\sin \tau,\sinh r\cos \phi,\sinh r \sin \phi)
\end{equation}
Denoting by $(r_i,\tau_i,\phi_i)$ the parameters of the corners of the null polygon, a convenient parameterization of such corners is
\begin{equation}
\phi_i=(i-1)\frac{\pi}{2}+\frac{\pi}{4},~~\tau_i=\frac{(-1)^{i+1}}{2}\arccos({\rm sech} R_S ~{\rm sech} R_T )
\end{equation}
where $r_1=r_3=R_S,~r_2=r_4=R_T.$
and 
\begin{equation}
S=4\sinh^2 R_S,~~T=4\sinh^2 R_T.
\end{equation}
If we project the polygon onto the $X^1,X^2$ plane we get a rhombus. We can nicely parametrize the polygon over that base using $(x,y)$ coordinates 

\begin{equation}\badat{3} \label{xy}
x&=\frac{\csch R_S+\csch R_T}{\sqrt{2}L}\sinh r \cos\phi+\frac{\csch R_S-\csch R_T}{\sqrt{2}L}\sinh r \sin \phi ,\\
y&= \frac{-\csch R_S+\csch R_T}{\sqrt{2}L}\sinh r \cos\phi+\frac{\csch R_S+\csch R_T}{\sqrt{2}L}\sinh r \sin \phi
\eadat\ee
which take values in the $[-1,1]^2$ square. (cusp $1$ is $(1,1)$, cusp $2$ is $(-1,1)$, edge $12$ is the horizontal line connecting these two, etc.) The minimal area problem becomes the problem of solving the equations of motion derived from the Nambu-Goto action for the height function $\tau(x,y)$ on the square $[-1,1]^2$ with appropriate boundary values. Pulling back the embedding space metric to~\eqref{eq:Xemb} gives
\be\badat{3}
&ds^2=-\frac{1}{4}L^2 (4 +  (x^2 + y^2) \sinh^2 R_S+(x-y)^2\sinh^2 R_T) d\tau^2 \\
&+\frac{L^2\sinh^2R_S\sinh^2R_T}{8(4 +  (x^2 + y^2) \sinh^2 R_S+(x-y)^2\sinh^2 R_T) } \times\bigg( 
(y^2+{\rm csch}^2 R_S+{\rm csch}^2 R_T)dx^2\\
&+(x^2+{\rm csch}^2 R_S+{\rm csch}^2 R_T)dy^2-2(xy+{\rm csch}^2 R_S-{\rm csch}^2 R_T)dxdy\bigg)
\eadat\ee
Now we want to further pull this back to our surface $\tau(x,y)$ in which case the induced metric has determinant
\be\badat{3}\label{eq:ext}
h&=\det h_{ab} \\ 
& = \frac{L^4\sinh^2 R_S \sinh^2 R_T}{4}\bigg[\frac{4}{(4+(x+y)^2\sinh^2 R_S+(x-y)^2\sinh^2 R_T)} 
\\
&-[(y^2+{\rm csch}^2 R_S+{\rm csch}^2 R_T)(\p_y\tau)^2+(x^2+{\rm csch}^2 R_S+{\rm csch}^2 R_T)(\p_x\tau)^2\\
&+2(xy+{\rm csch}^2 R_S-{\rm csch}^2 R_T) \p_x\tau \p_y \tau] \bigg ].
\eadat \ee
The equation of motion is then 
\be
0=\p_i\left[\frac{1}{\sqrt{h}}\frac{\delta h}{\delta \p_i \tau}\right]
\ee
where we've used that $\delta h/\delta\tau =0$. The equation of motion might seem a bit messy in full generality but it is actually not so bad; in particular it only involves up to second order derivatives in $\tau$ so it it not that much more complicated than a Laplace equation say. We find
\be\badat{3}
0 & = \p_x^2\tau [-8(4(S+T)+STx^2)+2(16+T(x-y)^2+S(x+y)^2)^2(\p_y\tau)^2]\\
&+\p_y^2\tau [-8(4(S+T)+STy^2)+2(16+T(x-y)^2+S(x+y)^2)^2(\p_x\tau)^2]\\
&+(\p_x\tau)^3x(4(S+T)+STx^2)(16+T(x-y)^2+S(x+y)^2)\\
&+(\p_y\tau)^3y(4(S+T)+STy^2)(16+T(x-y)^2+S(x+y)^2) \\
&+(\p_x\tau)^2 \p_y\tau(4(S+T)x+8(T-S)y+3ST xy^2)(16+T(x-y)^2+S(x+y)^2)\\
&+(\p_y\tau)^2 \p_x\tau(4(S+T)y+8(T-S)x+3ST yx^2)(16+T(x-y)^2+S(x+y)^2)\\
&-4\p_x\p_y\tau(16(T-S)+4STxy+(16+T(x-y)^2+S(x+y)^2)^2\p_x\tau \p_y\tau)\\
&-32ST(y\p_y\tau+x\p_x\tau).
\eadat\ee
The boundary conditions, on the other hand, are very simple. They can be summarized into the condition
\begin{equation}\label{eq:bndy}
\begin{split}\tau(x,y)|_{\p ([-1,1]^2)}&=\arctan\left(\frac{\sqrt{4+S}-\sqrt{4+T}+(\sqrt{4+S}+\sqrt{4+T})xy}{\sqrt{4+S}+\sqrt{4+T}+(\sqrt{4+S}-\sqrt{4+T})xy}\right. \\
&\hspace{3cm}\left. \times \tan\left(\frac{1}{2}\arccos\frac{4}{\sqrt{4+S}{\sqrt{4+T}}}\right)\right) \,. 
\end{split}
\end{equation}

\subsection{Numerics for $S=T$}

We now specialize the boundary value problem above to the one-parameter family
\begin{equation}
    S=T=4\sinh^2 R.
\end{equation}
In this case the boundary condition simplifies to
\begin{equation}
    \tau(x,y)\big|_{\partial[-1,1]^2}
    =
    \arctan\left[
    xy\,\tan\left(
    \frac12\arccos(\operatorname{sech}^2 R)
    \right)
    \right].
\end{equation}
Thus, for each value of $R$, the numerical problem is to solve the Nambu-Goto equation of motion for a single function $\tau(x,y)$ on the square, with the above boundary values.\footnote{The computation was done by discretizing the square using a Chebyshev  grid. The boundary values were imposed exactly and then we used
continuation in $R$: starting from small $R$, each accepted solution was used as the initial seed for the next value of $R$. (The larger $R$ is the harder the numerics get. This is perhaps expected as the surface is becoming bigger and bigger with nearly null cusps that need to be resolved close to the boundary for very large $R$.) At each step we relaxed the discretized equations of motion and monitored both the equation-of-motion residual and the imaginary part of the Nambu-Goto area. Only points for which the integrated imaginary part of the area remained below the percent level were included in the plots below. (The imaginary part -- when present -- is of course pure numerical error from the surface become timelike for a bit close to the boundary, most notably close to the cusps.)}

In the end, we found the numerical surface for $0\leq R \leq 2.84$. The resulting numerical area is shown in Figure~\ref{fig:area-numerics}.
The left panel shows the Nambu-Goto area as a function of $R$, while the right panel shows the same data divided by $R^2$, which should interpolate between a constant at weak coupling and a different constant at strong coupling according to the Gross-Manes and Alday-Maldacena predictions. 
\begin{figure}
    \centering
    \includegraphics[width=\linewidth]{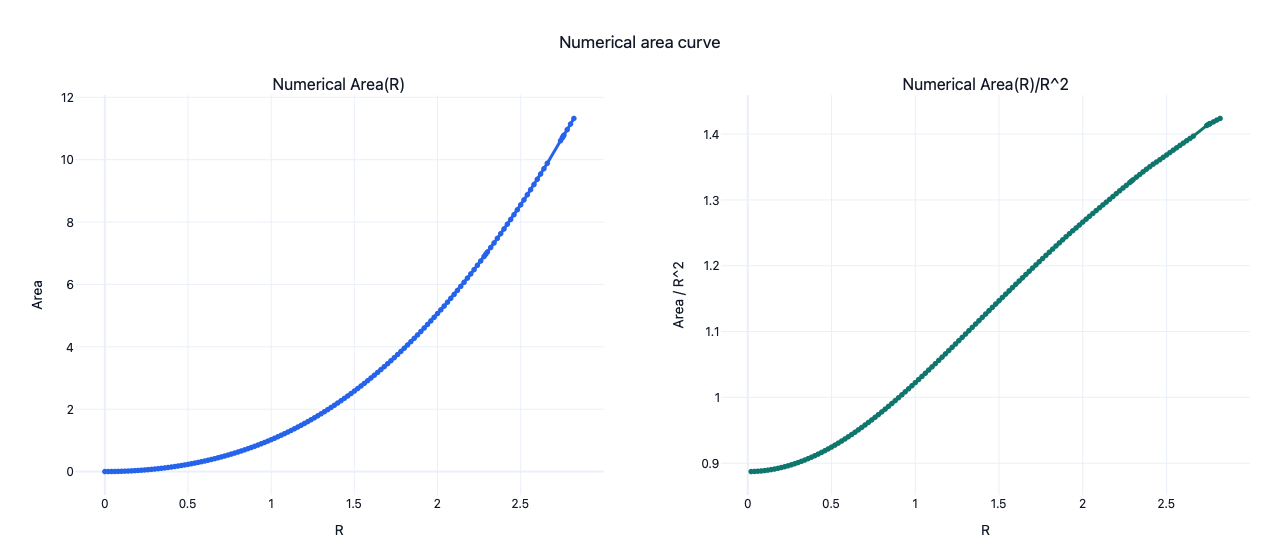}
    \caption{Numerical Nambu-Goto area for the symmetric kinematics $S=T=4\sinh^2 R$. The left panel shows the real part of the area computed from the relaxed surface.  The right panel shows $\texttt{Area}(R)/R^2$, which makes the interpolation between the small- and large-invariant regimes more visible.  Only numerical points passing the equation-of-motion and integrated-area reality checks are shown.}
    \label{fig:area-numerics}
\end{figure}

\subsection{Matching Gross-Manes for $S,T\ll 1$} \label{GMwithNGsec}

At small $S$ and $T$ we can easilly solve the equations of motion above. Alternatively, we can find the solution by translating the Gross-Manes solution to these variables as explained in appendix \ref{GMNGAp}. Either way, we find, in the limit $S=T \ll 1$, 
\begin{equation}
    \tau_{\rm GM}(x,y) = \frac{\sqrt{S}}{\sqrt{2}\pi}\arcsin\left[\sin\left(\frac{\pi x}{2}\right)\sin\left(\frac{\pi y}{2}\right)\right] \label{GMSequalT}
\end{equation}
The leading area is the Gross-Manes result specialized to $S  =T$. Our numerics can be seen to reproduce this solution and the associated area. In fact, accounting for the first correction as well we have the form
\begin{equation}\label{NGGMArea}
    \texttt{Area}(T,T) \simeq \frac{T}{2\pi}\log 4 + \gamma T^2
\end{equation}
with the constant $\gamma$ found to be
\begin{equation}
    \gamma = \frac{\pi^2 \log(16)-45\zeta(3)}{96\pi^3}.
\end{equation}
This same constant can be reproduced by the analytic leading correction in the expansion of the surface around flat space as we are going to describe in Section \ref{sec:flat-space-corrections}. Moreover, as we will also show it is compatible with the AdS amplitude analysis from \cite{Alday:2025pmg}. The plots in Figure \ref{fig:small-r-plots} show the numerical area compared with the leading Gross-Manes and leading Gross-Manes plus first curvature correction.
\begin{center}
    \begin{figure}[t]
        \centering
        \includegraphics[width=1\linewidth]{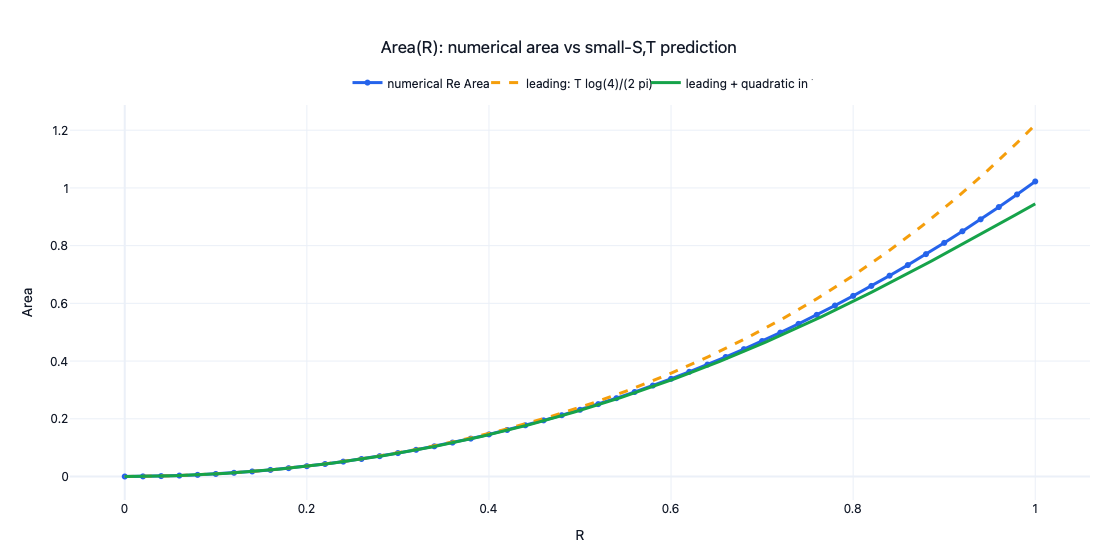}
        \caption{In blue we have the numerical area extracted from the Nambu-Goto analysis. In yellow the leading Gross-Manes result and in green the leading Gross-Manes plus first AdS radius correction.}
        \label{fig:small-r-plots}
        \label{fig:gross-manes-correction}
    \end{figure}
\end{center}

\subsection{Matching Alday-Maldacena for $1\ll S,T$}

In the Alday-Maldacena, $S,T\gg 1$ regime, we can identify in the gauge we are working now that the solution is given by
\begin{equation}
    \sin 2\tau_{\rm AM}(x,y)=\frac{S(x+y)^2-T(x-y)^2}{16+T(x-y)^2+S(x+y)^2}.    
\end{equation}
We can then write the general expected expansion of the area at large $S$ and $T$ as
\begin{equation}
    \texttt{Area}(S,T) = \frac{1}{2} \log S\log T + \beta (\log S + \log T) + C + \cdots,\quad \text{as $S,T\to\infty$},
\end{equation}
where $\beta$ is the constant that has been fixed by the Regge limit large $T$ correction. Then for the special kinematics $S = T$ we have
\begin{equation}
    \texttt{Area}(T,T) = \frac{1}{2} (\log T)^2 + 2\beta\log T + C + \cdots,\quad \text{as $T\to \infty$}
\end{equation}
In that case, if we look at the large $T$ expansion of the area we got from numerics, we must have the behavior
\begin{equation}
    \lim_{S\to \infty}\left(\texttt{Area} - \tfrac{1}{2} (\log T)^2 - 2 \beta \log T\right) = C.
\end{equation}
To check this behavior we plot in Figure~\ref{fig:large-R-constant} the subtraction leading and leading plus subleading behavior from the numerical area. When we subtract both corrections we can see visually that a constant is being approached as expected. Since we were unable to attain larger values of $R$, however, it is difficult to find the value of $C$ with high precision, but nevertheless from the current numerics we are able to infer $C \simeq 1.45$.
\begin{center}
    \begin{figure}[t]
        \centering
        \includegraphics[width=\linewidth]{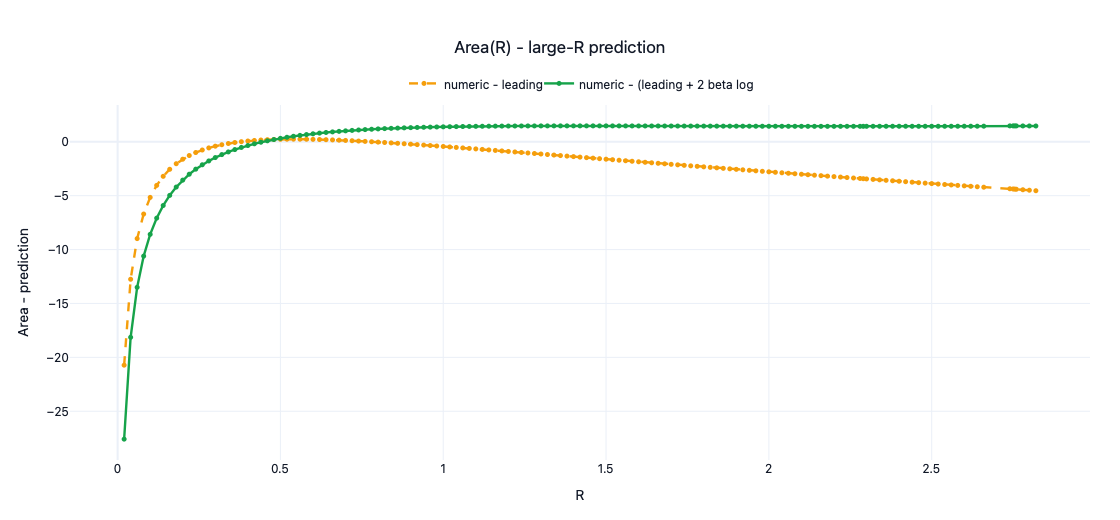}
        \caption{The difference between the numerical $\texttt{Area}(R)$ curve and the predictions. In yellow we just used the leading prediction $\tfrac{1}{2}(\log T)^2$, and in green we used the Regge corrected one $\tfrac{1}{2}(\log(T))^2 + 2\beta \log T$. In the latter case a constant is seem to be approached as expected. }
        \label{fig:large-R-constant}
    \end{figure}
\end{center}

\section{Flat Space (and Flat Space Corrections)} \label{sec:flat-space-corrections}

\subsection{Conformal Gauge: Reviewing Flat Space in the Polyakov Formalism} \la{sec:flatlimit}

In this section we will explore the flat space minimal surfaces and corrections around AdS. To do so we must start by reviewing the canonical result of Gross and Manes~\cite{Gross:1989ge}. Nothing in this subsection is new; we include this review material here for completeness and to set the notation for the next more novel subsections. Although the discussion begins with string amplitudes, it will shortly lead to the polygonal minimal surfaces of interest. 

In 1987, Gross and Mende studied high energy closed string scattering at fixed angle. In other words, $s$ and $t$ are both large \cite{Gross:1987ar}. In this high energy limit the string vertex operators
\beq
V_a(z,\bar z) \simeq \exp(i\, k_a \cdot x) \times \texttt{unimportant non-universal prefactor}
\eeq
In this section $x$ is a $d$ dimensional vector in $\mathbb{R}^{1,d-1}$. Gross and Mende pointed out that the insertion of these vertex operators with large momenta source a classical string. The amplitude given by a full path integral is dominated by a saddle point solution of the form 
\beq
x = \frac{1}{\pi}\sum_{a=1}^n k_a  \log |z-\sigma_a|\label{GM} \,.
\eeq 
If $\sigma_a$ is real, the solution with $z$ in the full complex plane also describes a nice open string amplitude -- relevant for high energy scattering of open strings -- as long as we restrict $z$ to be in the upper half plane. That open string analysis was carried out by Gross and Manes two years later in 1989. 

Solution (\ref{GM}) clearly has the right boundary conditions close to each puncture: 
\beq
x\simeq \frac{k_a}{\pi} \tau \,, \qquad z \simeq \sigma_a + e^{\tau \pm i \sigma} \,, \qquad \tau \to -\infty \label{GMcusp}
\eeq
It also clearly obeys the equations of motion 
\beq
\partial \bar \partial x=0
\eeq
since $x$ is manifestly harmonic. To claim it is a minimal surface, however, we also need to impose the Virasoro constraints 
\beq
0=\partial x \cdot \partial x 
\eeq
which require the puncture locations to obey a nice set equations resembling a sort of electrostatic equilibrium condition, 
\beq
0=\sum_{b\neq a} \frac{k_a\cdot k_b}{\sigma_{a}-\sigma_b} \,, \qquad a=1,\dots,n \label{SEs}
\eeq
together with the nullness condition $k_a^2=0$.\footnote{At high energy, even if we were scattering massive particles, we could effectively drop $k_a^2$ as it would be highly suppressed compared to $k_a\cdot k_b$; this makes this high energy limit extremely universal and ubiquitous in string theory.
}
Today these equations (\ref{SEs}) are usually called the \textit{scattering equations} or the CHY equations. They reappeared in 2013 in the seminal work of Cachazo, He and Ye \cite{Cachazo:2013hca}. Curiously, they reemerged in the extreme opposite limit to the one we have been discussing, namely in the low energy field theory limit. In that limit, we now know that tree level scattering amplitudes for a vast myriad of theories, from pions to Yang-Mills and gravity, can be cast as a sum over the solutions to these equations \cite{Cachazo:2013iea,Cachazo:2013gna,Cachazo:2014nsa,Cachazo:2014xea}. 

At first sight, there is a difference between how scattering equations show up at low and high energy. At low energy, scattering amplitudes are given by expressions involving summing over all scattering equation solutions. At high energy, string scattering amplitudes are dominated by classical saddles and thus, often, we only care about the dominant one.\footnote{This difference is probably superficial. Probably a better way to think of the string path integral is to decompose it into a sort of sum of several regions, thimbles of sorts, each containing one of the saddles. In the high energy limit one is much bigger than the other ones but they are always there. Some works in this direction include \cite{Mizera:2017cqs,Mizera:2017rqa}. 
 }

A nice observation by McGreevy and Sever is that, \textit{if} (\ref{GM}) is a good minimal solution, so is the $T$-dual solution obtained by replacing $\log z \bar z$ by $\log z/\bar z$ therein, 
\beq
x = \sum_{a=1}^n k_a \,\mathcal{L}_a(z,\bar z) \,,\qquad \mathcal{L}_a(z,\bar z) \equiv \frac{1}{2\pi i} \Big(\log\Big(\! -\frac{z-\sigma_a}{\bar z-\sigma_a}\Big) - i \pi \Big) \label{xPolygon}
\eeq 
After all, the solution is again clearly harmonic and the very same scattering equations ensure it is a proper extremal surface.

This solution describes a nice extremal null polygonal surface. Indeed, for $z,\bar z=x\pm i 0$ we get $\mathcal{L}_a(z,\bar z) \to \theta(x-\sigma_a)$ so that the solution has a very nice behavior in the real line: in the edges $\sigma_a <x<\sigma_{a+1}$ between the punctures the solution $x$ is fixed; then, as we pass by $\sigma_a$ is jumps by $k_a$. \footnote{Indeed, with this new solution, the analogue of (\ref{GMcusp}) would be 
\beq
x\simeq \sum_{b<a} k_b+ k_a \frac{\pi-\sigma}{\pi}  \,, \qquad z \simeq \sigma_a + e^{\tau \pm i \sigma} \,, \qquad \tau \to -\infty \,, \qquad \sigma \in [\pi,0]
\eeq
so that as we go around the puncture $\sigma_a$ drawing a small half-circle in the upper half plane, the solution jumps from the cusp $X_a=\sum_{b<a} k_b$ to the cusp $X_{a+1}=\sum_{b\le a} k_b$.} This is an important and a bit unusual feature of this solution: worldsheet punctures are mapped to polygon edges while worldsheet edges (segments between the punctures) are mapped to the polygon cusps. (This is of course very different from all our polygon description above based on the Nambu-Goto solution where no such $\texttt{points} \leftrightarrow \texttt{lines}$ swap took place.) A simple consequence of this point is that there is no contribution to the area in the real line in the segments between the punctures since each of those segments correspond to a single point in target space. 

Note that the original and T-dual solution have the same area\footnote{
There are simple factors of $2\pi$ and string tension floating around when relating areas and actions and we sometimes implicitly omit in intermediate expressions. We restore them all at the end in important expressions; they are crucial, of course, when plotting all quantities and matching various asymptotics as illustrated in the previous section.} given by 
\beq
\texttt{Area}=\int d^2z\, \partial x \cdot \bar \partial x = \sum_{a \neq b} \frac{k_a \cdot k_b}{4\pi^2} \int \frac{d^2 z}{(z-\sigma_a)(\bar z-\sigma_b)} \,.
\eeq
(The property that we alluded to above -- namely that in the real line the area integrand should vanish -- is easy to see: For $\bar z=z$ the integrand is indeed zero since it reduces precisely to the Virasoro conditions!)

At this point we see that there are some null polygon solutions which are given by high energy string solutions up to some simple T-duality transformation. Are \textit{all} null polygons in flat space described by these GM polygon solutions (\ref{xPolygon})? If the polygon is space-like -- i.e. if no two points along the polygon boundary are time-like separated -- then we can prove a simple neat theorem that says \textit{yes!} As far as we can tell this is a new result and we will thus postpone it to the next section and continue with our review here. 

Evaluating the area is a fun sequence of simple steps:
\beqa
\texttt{Area}&=&\int\limits_\text{UHP} d^2 z\, \partial  {x} \cdot \bar \partial {x} \nn  \\
&=&\frac{1}{4\pi^2} \int\limits_\text{UHP} d^2 z\, \sum_{a,b}  \frac{k_a}{z-\sigma_a} \cdot \frac{k_b}{\bar z-\sigma_b} \nn  \\
&=&\frac{1}{4\pi^2}\sum_{a,b} k_a\cdot k_b\int\limits_{0}^\infty dy \int\limits_{-\infty}^\infty dx   \frac{1}{(x+i y-\sigma_b)(x-i y-\sigma_a)} \nn  \\
&=&\frac{i}{2\pi}\sum_{a,b} k_a\cdot k_b\int\limits_{0}^\infty dy    \frac{1}{(2i y+\sigma_a-\sigma_b)} \nn \\
&=&\frac{1}{4\pi}\sum_{a\neq b} k_a\cdot k_b \log(\sigma_a-\sigma_b)  \label{final}
\eeqa
To get the last equality we note that we can replace $\infty$ by a large $\Lambda$ which drops out when using momentum conservation together with $k_a^2=0$. 

We see that the scattering equations can alternatively be derived by extremizing the final expression (\ref{final}) which we can conveniently think of as a sort of electrostatic potential. 

For four points the scattering equations have a single solution given by 
\beq
\frac{(\sigma_1-\sigma_2)(\sigma_3-\sigma_4)}{(\sigma_1-\sigma_3)(\sigma_2-\sigma_4)} = \frac{S}{T} = \frac{k_1\cdot k_2}{k_1\cdot k_3}
\eeq
and the area (\ref{final}) simply 
\beq
-2\pi \,\texttt{Area}=S \log S+T \log T-(S+T) \log (T+S) \label{final4} \,.
\eeq
in perfect agreement with the leading flat space behavior of the expression (\ref{NGGMArea}) we derived above, in the $S = T$ case kinematics, from the Nambu-Goto action. For $S\gg T$ this becomes  
\beq
2\pi \,\texttt{Area} \simeq  T \log S  \label{final4} \,.
\eeq
in perfect agreement with the general Regge expectation discussed before, see (\ref{ReggeGM}). 

As we said at the beginning, this is all well known. What follows are some observations related to this problem which we have not encountered elsewhere. 

\subsection{Some New Observations}

Now that we have reviewed the Gross Manes result in the previous subsection, let us turn to some new observations about the flat space solution. First, we will establish the existence of solutions to the scattering equations for kinematics that leave the polygon spacelike. Then we will show how to use these solutions to parameterize the surface in embedding space as a rational equation. Finally, we use this pole structure to examine the fate of applying integrability techniques to this solution. 

\subsubsection{One Solution for each Space-like Polygon}
Can we use (\ref{xPolygon}) for any space-like polygon? Yes. Indeed: 

\begin{theorem}Consider a closed null polygon with ordered edges $(p_1,...,p_n)$ and let 
\begin{equation}
X_{i, j} \equiv\left(p_i+p_{i+1}+\cdots+p_{j-1}\right)^2 
\text { for } 1 \leq i<j \leq n .
\end{equation}
The condition $X_{i,j}>0$ for all chords is sufficient to guarantee that there is a real solution to the scattering equations with the correct ordering of the punctures. 
\end{theorem}
\begin{proof} As in~\cite{
Arkani-Hamed:2017mur} we will start by gauge fixing our punctures so that
\be
\sigma_1=0,~~\sigma_{n-1}=1,~~\sigma_n=\infty
\ee
Because we have a fixed polygon we are interested in a particular planar ordering where
\be\label{dom}
0\le \sigma_{2}\le \sigma_3\le...\le \sigma_{n-2}\le 1.
\ee
Let us call this ordered region of configuration space  $C\subset [0,1]^{n-3}$. 
As in~\cite{Cachazo:2016ror} we will use the fact that the scattering equations
\be
\sum_{\overset{i=1}{i\neq j}}^n\frac{s_{ij}}{\sigma_{i}-\sigma_j}=0
\ee
can be viewed as extrema of the potential
\begin{equation}
V(\sigma)=-\sum_{2 \leq a<b \leq n-2} s_{a b} \log \left|\sigma_a-\sigma_b\right|-\sum_{a=2}^{n-2} s_{a 1} \log \left|\sigma_a\right|-\sum_{a=2}^{n-2} s_{a n-1} \log \left|1-\sigma_a\right|.
\end{equation}
We claim that within the region $C$, the potential is bounded from below if the conditions in the theorem are met, and that the desired real solution is $\sigma_a={\rm argmin}_C V$.

To see this, first note that the potential only diverges when some subset of punctures collide. This means that the potential is finite in the interior and will blow up on $\p C$. Since we are restricted to a planar ordering, higher codimension components of $\p C$ involve sets of {\it consecutive} punctures colliding. Using the telescoping identity
\be\label{tele}
\sum_{I\le i<j\le J} s_{ij}=X_{I,J+1}
\ee
we see that if punctures $I$ through $J$ collapse so that $\sigma_{i}-\sigma_{j}\sim\epsilon \ll 1$ the potential will diverge as
\be
V\sim -X_{I,J+1}\log|\epsilon|.
\ee
This is positive whenever the chords $X_{i,j}$ are positive. If multiple subsets of punctures collide this property continues to hold.  
Since $C$ is compact and $V$ is lower semi-continuous $V$ achieves its minimum on $C$. Since $V$ diverges on $\p C$ this minimum will be obtained in the interior. Since $V$ is continuous in the interior this minimum will be a local minimum and thus $\sigma_a={\rm argmin}_C V$ is a solution to the scattering equations.
\end{proof}

\subsubsection{Gross-Manes without Worldsheet}
We just found our minimal surfaces in flat space in parametric form in the so-called conformal gauge where $x=x(z,\bar z)$ is given by (\ref{xPolygon}). Can we describe these surfaces in a more invariant way without commiting to a particular set of coordinates? Yes. Indeed: 

Given a polygon with $n$ edges with momenta $k_{a=1,\dots,n}$, consider a set of dual vectors~$q_{a=1,\dots,n-1}$ defined by
\beq
q_a \cdot k_b = \delta_{a,b} \,, \qquad a,b=1,\dots,n-1
\eeq
Note that we did not (need to) use the last momentum $k_n=-k_1-\dots-k_{n-1}$. For these $q_a$ momenta to make sense without any subtlety let us assume that we are in $\mathbb{R}^{1,d-1}$ with
\beq
d\ge n-1  \,.
\eeq
If we now dot (\ref{xPolygon}) with $q_a$ we find $n-1$ relations:
\beq
Q_a\equiv \exp(2\pi i \,q_a \cdot x)=\frac{z-\sigma_a}{\bar z-\sigma_a} \Big/ \frac{z-\sigma_{n}}{\bar z-\sigma_{n}}
\label{Qvar}
\eeq
If we use two of these relations to solve for $z(Q_a,Q_b)$ and $\bar z(Q_a,Q_b)$ then the remaining relations are the surface we are after where the worldsheet coordinates are gone as promised. In the end, when doing this the final relations we get between the $Q_a$'s take the beautiful form
\beq
\frac{(Q_a-Q_b)(Q_c-Q_d)}{(Q_a-Q_c)(Q_b-Q_d)}=\frac{(\sigma_a-\sigma_b)(\sigma_c-\sigma_d)}{(\sigma_a-\sigma_c)(\sigma_b-\sigma_d)}=\texttt{constant}_{abcd} \label{finalQ}
\eeq
where $a,b,c,d=1\dots,n$ and where the last $Q_n$ is defined as $Q_n=1$ consistently with \eqref{Qvar}.
We see that the shape of the solution can be summarized as the statement that cross-ratio of the surface coordiates $Q_a$ are constant along the surface! The value of these constants $\texttt{constant}_{abcd}$ is then fixed by imposing that this surface is minimal. The answer is that these constants are given by cross-ratios of $\sigma$'s which obey scattering equations. 

\begin{figure}[t]
    \centering
\includegraphics[width=1\textwidth,angle=0,trim={0cm 0cm 0cm 0cm}, clip=true]{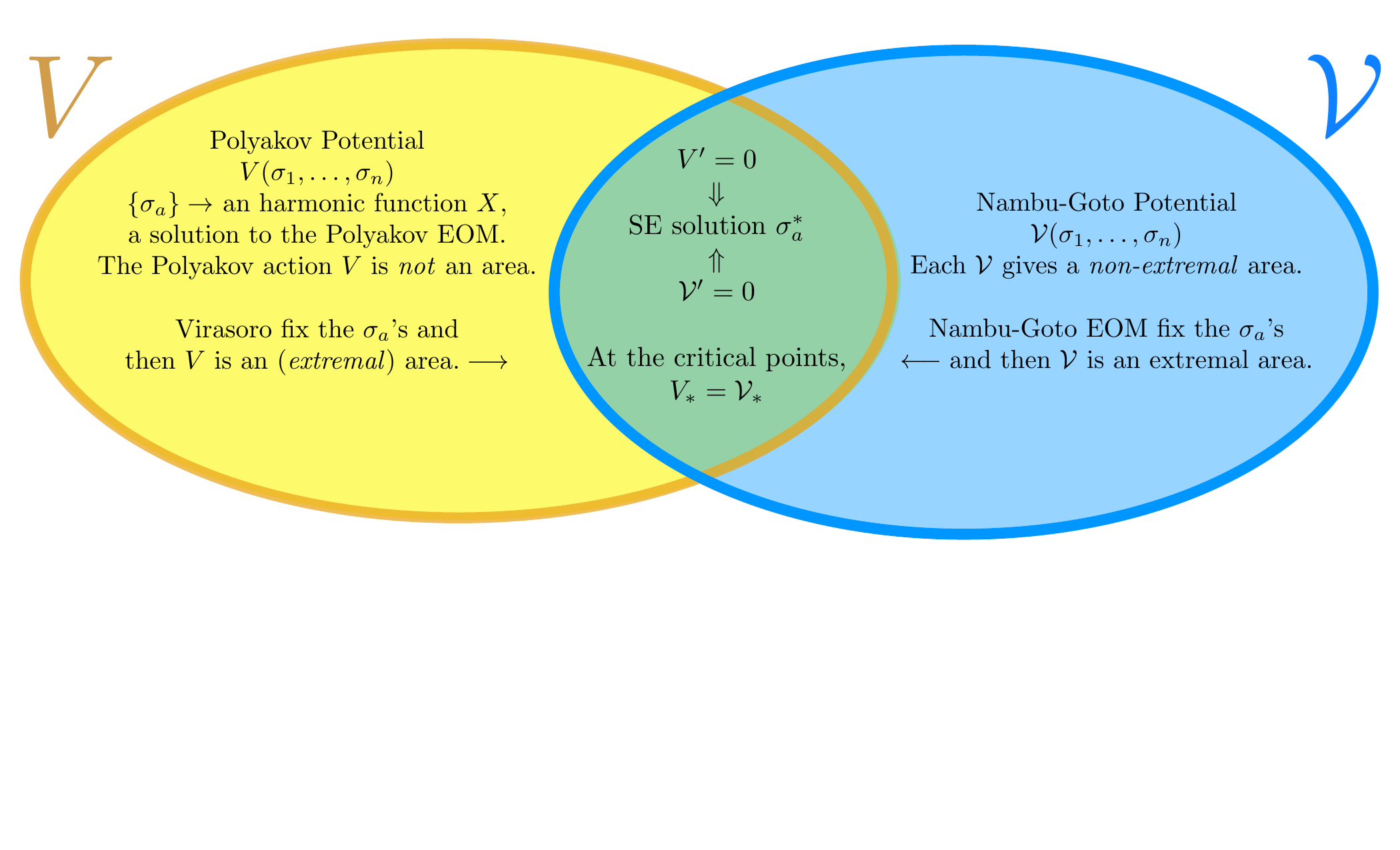}
\vspace{-4cm}
    \caption{Off-shell, the Nambu-Goto and the Polyakov potentials are different but are extremized by the same solutions to the scattering equations and have the same value at these critical points. Off-shell, the Nambu-Goto potential is always an area but not an extremal one while the Polyakov potential is always a solution to the EOM (an harmonic function) but is not an area.}
        \label{fig:potentials}
\end{figure}

In appendix \ref{GMNGAp} we verify that the relation (\ref{finalQ}) for four points leads to an explicit solution for $Q_3(Q_1,Q_2)$ which precisely matches the Nambu-Goto solution (\ref{GMSequalT}) anticipated in section \ref{GMwithNGsec}.

It is interesting to compare the Nambu-Goto and the Polyakov functionals, see figure~\ref{fig:potentials}.\footnote{We thank Nima Arkani-Hamed and Carolina Figueiredo for interesting discussions and correspondence on this particular point.} Both (\ref{xPolygon}) and (\ref{finalQ}) describe solutions ending on null polygons but only when the $\sigma_a$ obey the scattering equations are these solutions the same; othewise they describe two different off-shell deformations of the minimal solution and would be nice to see if both could be useful. At the quantum level, for instance, can we learn something from these functionals?

\subsubsection{Three Dimensions and some Integrability}
Here we unveil some integrability structures in the flat space analysis. Of course, if all we wanted were the classical polygons in flat space, this would be quite an overkill since we already got those -- and their areas -- by two different methods (Polyakov and Nambu-Goto). The goal here is, rather, to unveil structures that we can hopefully lift to $AdS$ in the near future.   

We will focus for a while on the case where the minimal surface is embedded in three-dimensional space-time $\mathbb{R}^{1,2}$. This is totally general if we are describing null squares with~$n=4$ since we can always map a general null polygon to a three-dimensional subspace of~$\mathbb{R}^{1,d}$ without any loss of generality. For~$n>4$, however, the restriction to three dimensional kinematics is a simplifying assumption. 

\subsubsection*{The holomorphic function $p(z)$}
Let us construct the complexified\footnote{The real unit normal is $$N_R^a = \frac{i\epsilon^a{}_{bc}\partial x^b\bar\partial x^c}{\partial x\cdot\bar\partial x}$$ obeying $N_R^2 =-1$. It is related to $N$ by $N = \pm i N_R$ with the sign depending on orientation.} unit normal vector ${N}$ to a minimal surface in three dimensional flat space. It is defined as the unit vector orthogonal to $\partial x$ and $\bar \partial x$ and normalized by $N\cdot N = 1$;  since we are in three dimensions this defines $N$ uniquely up to an overall sign. So if we guess $N$ we are done. A good guess is 
\beq
{N}=a_1\, \partial {x}+ a_2\, \partial^2 {x} 
\eeq
The coefficients can be fixed by (a) imposing orthogonality to the tangent vectors to the surface and by (b) normalizing $N$. We get\footnote{Note that this vector is automatically orthogonal to $\partial x$ since Virasoro gives $\partial {x} \cdot \partial {x}=0$ but also, taking a derivative of this equation, $\partial {x} \cdot \partial^2 {x}=0$. So we simply need to impose orthogonality with $\bar \partial \vec{x}$ to fix the ratio of $a_2/a_1$. The overall normalization is then fixed by normalizing $N$. Note that -- again because of Virasoro -- we have ${N}^2=a_2^2 \, \partial^2 {x}\cdot  \partial^2 {x}$ so the normalization condition is trivial and simply sets $a_2=1/\sqrt{\partial^2 {x}\cdot  \partial^2 {x}}$ so that at the end of the day we obtain (\ref{finalN})}
\beq
{N}=\frac{1}{\sqrt{\partial^2 {x}\cdot  \partial^2 {x}}} \Big(  \partial^2 {x}-\frac{ \partial^2 {x}\cdot  \bar\partial {x}}{ \partial {x}\cdot  \bar\partial {x}}\, \partial {x} \Big) \label{finalN}
\eeq
Of course, we can also write an equivalent expression for $N$ by swapping $\partial$'s and $\bar \partial$'s in this expression. 

With this normal vector we define two very simple and important functions
\beq
p(z) \equiv {N} \cdot \partial^2 {x} \,, \qquad \bar{p}(\bar z) \equiv {N} \cdot \bar\partial^2 {x}
\eeq
which are simple holomorphic and anti-holomorphic functions as anticipated by the left hand sides. Plugging ${N}$ in these expressions\footnote{This is very easy to establish using the condition $\partial x \cdot \partial^2 x=0$ -- which we referred to in the previous footnote 
}
leads to
\beq
p(z)=\sqrt{\partial^2 {x}\cdot  \partial^2 {x}} \label{pDef2}
\eeq
and similarly for $\bar p$. That this is holomorphic is now obvious: It suffices to apply $\bar\partial$ to this expression and use the equations of motion $\bar \partial \partial x=0$ (and thus $\bar\partial\partial^2 x=0$ as well). This representation is very compact but has this annoying square root which is actually not there for the Gross-Manes surface in three dimensions as we will see below. 

For the Gross-Manes solution (\ref{xPolygon}) we have
\beq
p(z)^2=\frac{-1}{4\pi^2}  \sum_{a \neq b} \frac{k_a\cdot k_b}{(z-\sigma_a)^2(z-\sigma_b)^2}  \la{sqrtP} \,.
\eeq
Naively this would decay at $1/z^4$ at large $z$ but actually it decays much faster, as $1/z^8$. This can be seen by expanding the right hand side at large $z$ and noting that the first four terms vanish either trivially\footnote{For example: for any function of $\sigma_a$ we have~$\sum_{a \neq b}  k_a\cdot k_b f(\sigma_a)=0$ since the sum over $b$ replaces $k_b$ by $-k_a$ by momentum conservation and then $p_a^2=0$.} or as a consequence of the scattering equations (i.e. the Virasoro condition)\footnote{Indeed, consider the expansion of the Virasoro constraints $0=\partial x\cdot \partial x$ at large z,  
\beq
0=\sum_{a\neq b} \frac{k_a \cdot k_b}{(z-\sigma_a)(z-\sigma_b)}=\frac{1}{z^2} \sum_{a\neq b} k_a \cdot k_b+\frac{2}{z^3} \sum_{a\neq b} k_a \cdot k_b \,\sigma_a+\frac{1}{z^4} \sum_{a\neq b} k_a \cdot k_b(2\sigma_a^2+\sigma_a \sigma_b)+\dots \label{expansionVir}
\eeq
Most terms in the right hand side are trivially zero -- see previous footnote. Some, however, lead to interesting sum rules such as \beq
\sum_{b\neq a} (k_a \cdot k_b)( \sigma_a \sigma_b) = 0 \,, \qquad \sum_{b\neq a} (k_a \cdot k_b) (\sigma_a^2 \sigma_b) = 0 \,. 
\eeq which similarly imply that the large $z$ expansion of (\ref{sqrtP}) only starts at $1/z^8$. These two conditions follow from the SE since these are equivalent to Virasoro. All in all we conclude that $p(z)^2\simeq B/z^8$ at large $z$ with  $B=\big(\sum_{a} k_a \sigma_a^2\big)\cdot \big(\sum_{b} k_b \sigma_b^2\big)=\sum_{a\neq b} (k_a\cdot k_b)(\sigma_a^2\sigma_b^2)$. As we will see below, this fixes $p(z)$ completely for $n=4$ and leaves $n-4$ conditions to fix for $n>4$. \label{footB}
}.
This means that 
\beq
p(z)=\frac{\texttt{Degree $n-4$ polynomial in $z$}}{\prod\limits_{a=1}^n (z-\sigma_a)}  \label{polynomial}
\eeq
Note that we wrote the numerator as a polynomial of degree $n-4$ rather than the square root of a polynomial of degree $2n-8$ as one would naively guess by taking the square root of (\ref{sqrtP}). The reason is that $p(z)$ and $\bar p(\bar z)$ are rational functions in three dimensions.\footnote{In higher dimensions, we could have defined the holomorphic quantity $p(z)$ using (\ref{pDef2}) without the need to introduce $N$ at all. But we could not use the argument that follows and therefore for $d>3$ the numerator of (\ref{polynomial}) should be replaced by the square root of a polynomial of degree $2n-8$.
} 
To see this we can write down $1=N\cdot N$ where in the right hand side we use the representation (\ref{finalN}) for $N$ with $\partial$ and also its other representation using $\bar \partial$. This gives 
\beq
1=\frac{1}{p(z) \bar p(\bar z)} \Big(\partial^2 {x}\cdot {\bar\partial}^2 {x}-\frac{( \partial^2 {x}\cdot  \bar\partial {x})( \bar\partial^2 {x}\cdot  \partial {x})}{ \partial {x}\cdot  \bar\partial {x}}\Big) \label{1equal}
\eeq
Clearly the parentheses is a rational function of $z$ and $\bar z$ when we plug (\ref{xPolygon}) in. It has no square roots and thus neither can $p(z)$ or $\bar p(\bar z)$. 

In particular, for $n=4$ the numerator in (\ref{polynomial}) is a simple constant which we can find by matching a residue of this expression with (\ref{sqrtP}) which tells us that near puncture $\sigma_a$ we have the expansion
\beq
2\pi \, p(z) \simeq \frac{1}{z-\sigma_a} \sqrt{-2\sum_{b\neq a} \frac{k_a\cdot k_b}{(\sigma_a-\sigma_b)^2}} \,. \label{residue}
\eeq 
Alternatively, we can use footnote \ref{footB}.

\subsubsection*{Liouville Equation and the Area Element}

Denote the area element $\partial x \cdot \bar \partial x$ as
\beq
e^\alpha \equiv \partial x \cdot \bar \partial x \,. \label{alphaDefFlat}
\eeq
Then evaluate $e^\alpha \partial \bar\partial \alpha$. Using Virasoro and the equations of motion we see that this quantity is nothing but the parentheses in (\ref{1equal}). Therefore we find that the function $\alpha$ obeys the modified Liouville equation 
\beq
\partial \bar\partial \alpha + p \bar p \,e^{-\alpha} = 0 \,. \la{LioEq}
\eeq
Liouville is of course a very well known integrable model. In a sense, we could call it uber-integrable as we can find its generic solution analytically: 
\beq
e^{\alpha(z,\bar z)}=-\frac{1}{2} \times p(z) \bar p(\bar z)  \times \frac{(g(z)-\bar g(\bar z))^2}{g'(z)\bar g'(\bar z)} \,.
\eeq
It solves (\ref{LioEq}) for any $g$ and $\bar g$. The solution we want corresponds to the most trivial possible choice, namely $g(z)=z$, $\bar g(\bar z)=\bar z$ so that 
\beq
\left. e^{\alpha(z,\bar z)}\right|_\texttt{Gross-Manes}=-\tfrac{1}{2}  p(z) \bar p(\bar z) (z-\bar z)^2\,. \label{alphapp}
\eeq
The map from the vector $x$ to an scalar $\alpha$ as in (\ref{alphaDefFlat}) is quite common in AdS studies~\cite{Hofman:2006xt,Alday:2009yn} and goes by the name of Pohlmeyer reduction~\cite{Pohlmeyer:1975nb,Grigoriev:2008jq}. (We will come back to it in the next section.) In flat space, we have not seen it used before.

Can we compute the area 
\beq
\texttt{Area}=\int d^2z \,e^{\alpha}
\eeq
using Integrability of Liouville theory? That could be very inspiring in guessing the AdS integrable structure. In previous integrability studies, see e.g. \cite{Alday:2010ub,Janik:2011bd,Kazama:2011cp,Caetano:2012ac}, a crucial step involved translating the area integrand into the product of two closed forms and then decomposing that two dimensional area integral into a sum of one dimensional cycle integrals of those forms using the so-called Riemann Bilinear identity. In appendix \ref{RBIAp} we explain how something similar can be done for the flat space area. Note that this two-to-one dimensional map is basically the usual KLT decomposition of closed string in terms of sums of products of open strings \cite{Kawai:1985xq}. For a very nice promising set of lectures on this direction see~\cite{Komatsu:2019xzz}, see also~\cite{Honda:2013pba}.

\subsection{Corrections around Gross-Manes}

In AdS, using embedding coordinates, we write
\beqa
X(z,\bar z)&=& X_\texttt{polygon}(z,\bar z)+ \delta X(z,\bar z)\\
X_\texttt{polygon}(z,\bar z)&=&\mathbb{X}_1+\sum_{j=1}^4 (\mathbb{X}_{i+1}-\mathbb{X}_i) \mathcal{L}_i \,, \qquad \mathcal{L}_j \equiv \frac{1}{2\pi i} \Big(\log\Big( -\frac{z-\sigma_j}{\bar z-\sigma_j}\Big) - i \pi \Big) \\ \label{Xpol}
\left(\begin{array}{c}
\mathbb{X}_1\\
\mathbb{X}_2\\
\mathbb{X}_3\\
\mathbb{X}_4 \end{array}\right)&=&\left(
\begin{array}{cccc}
 \frac{1}{2} \sqrt{4 L^2+\mathbb{S}} & 0 & \frac{\sqrt{\mathbb{S}}}{2} & 0 \\
 \frac{2 L}{\sqrt{\frac{\mathbb{S}}{L^2}+4}} & \frac{1}{2} \sqrt{\frac{4 L^2 \mathbb{S}}{4 L^2+\mathbb{S}}+\mathbb{T}} & 0 & \frac{\sqrt{\mathbb{T}}}{2} \\
 \frac{1}{2} \sqrt{4 L^2+\mathbb{S}} & 0 & -\frac{\sqrt{\mathbb{S}}}{2} & 0 \\
 \frac{2 L}{\sqrt{\frac{\mathbb{S}}{L^2}+4}} & \frac{1}{2} \sqrt{\frac{4 L^2 \mathbb{S}}{4 L^2+\mathbb{S}}+\mathbb{T}} & 0 & -\frac{\sqrt{\mathbb{T}}}{2} \\
\end{array}
\right) \nn
\eeqa
The simple results $(\mathbb{X}_i-\mathbb{X}_{i+1})^2=0$ and
\beqa
&&(\mathbb{X}_1-\mathbb{X}_3)^2=\mathbb{S},\quad(\mathbb{X}_2-\mathbb{X}_4)^2= \mathbb{T}
\eeqa
justify the use of $\mathbb{S}$ and $\mathbb{T}$ as our variables. Note that to compare to the variables $S$ and $T$ we have used in the previous we need to set $\mathbb{S} = L^2 S$ and $\mathbb{T} = L^2 T$. In that case the large $L$ expansion at fixed $\mathbb{S}$ and $\mathbb{T}$ here corresponds to the small $S$ and $T$ expansions in the previous sections. 

The frame defined here matches the frame used in the Nambu-Goto numerics up to an AdS$_3$ isometry and a permutation of the labelings of the corners of the polygon. More precisely, define
\begin{equation}\label{eq:rotation-frame-sec-3-to-4}
    \Lambda = \left(
\begin{array}{cccc}
 \cos (\theta ) & \sin (\theta ) & 0 & 0 \\
 \sin (\theta ) & -\cos (\theta ) & 0 & 0 \\
 0 & 0 & \frac{1}{\sqrt{2}} & \frac{1}{\sqrt{2}} \\
 0 & 0 & \frac{1}{\sqrt{2}} & -\frac{1}{\sqrt{2}} \\
\end{array}
\right),\quad \theta \equiv \frac{1}{2}\arccos\left(\operatorname{sech} R_S \operatorname{sech} R_T\right),
\end{equation}
with $S = 4\sinh^2 R_S$ and $T = 4\sinh^2 R_T$ as before. If $P_i$ are the cusps obtained by setting the corner values $(\tau,r,\phi) = (\tau_i,r_i,\phi_i)$ in \eqref{eq:Xemb}, we find that $\Lambda P_1 = \mathbb{X}_1$, $\Lambda P_2 = \mathbb{X}_4$, $\Lambda P_3 = \mathbb{X}_3$ and $\Lambda P_4 = \mathbb{X}_2$.

We solve the equations of motion perturbatively to find $\delta X$ in a $1/L$ expansion. Note that $X_\texttt{polygon}(z,\bar z)$ nicely sets correct boundary conditions for a null polygon in $AdS$ hence in the real line we should impose $\delta X(z,z)=0$. $\delta X$ contains a trivial $1/L$ term and an interesting $1/L^2$ term. The final expression is too messy to put here. It has transcendentally degree three polylogarithms and so on. This analysis is very similar to the one in \cite{Alday:2023pzu}. 
The final expression as well as checks of the formulae below can be found in the ancillary file \texttt{ForPaper.nb}. 

From the final expressions we can compute $\alpha=\log \partial X \cdot \bar\partial X$ and $p^2=\partial^2 X \cdot \partial^2 X$. Let us first remind ourselves that to leading order we can simply drop $\delta X$ and expand $X_\texttt{polygon}(z,\bar z)$ to leading order in $1/L$ to find 
\beq
p_0(z)\bar p_0(\bar z)= \dfrac{\mathbb{T}}{4\pi^2} \dfrac{(\sigma_1-\sigma_2)(\sigma_1-\sigma_3)(\sigma_2-\sigma_4)(\sigma_3-\sigma_4)}{ |z-\sigma_1|^2|z-\sigma_2|^2|z-\sigma_3|^2
|z-\sigma_4|^2}\, , \qquad e^{\alpha_0} = -  \tfrac{1}{2}\,p_0(z) p_0(\bar z) (z-\bar z)^2
\eeq
where Virasoro fixes the scattering equations which are equivalent to
\beq
\chi \equiv \frac{(\sigma_1-\sigma_2)(\sigma_3-\sigma_4)}{(\sigma_2-\sigma_3)(\sigma_1-\sigma_4)}= \frac{\mathbb{S}}{\mathbb{T}} \,.
\eeq
Next, the correction, which we write as
\beq
\alpha(z,\bar z)=\alpha_0(z,\bar z)+\frac{1}{L^2}\, \alpha_2(z,\bar z)+\dots \,, \qquad  p(z)=p_0(z)\Big(1+\frac{1}{L^2}p_2(z)+\dots\Big)
\eeq
Here is what we find from this analysis: 
\begin{itemize}
\item The scattering equations -- that is the constraint from Virasoro -- gets modified into\footnote{Note that we never need to solve any scattering equations since we can work parametrically: We simply fix $\mathbb{T}$ and the positions of the four puncture and we read off $\mathbb{S}$ from (\ref{Seq}). } 
\beq
 \frac{\mathbb{S}}{\mathbb{T}}=\chi\left(1+\frac{\mathbb{T}}{4\pi^2L^2}f(\chi)\right) \label{Seq}
\eeq
where 
\beq 
f(\chi)=6 (\chi +1) \text{Li}_2(-\chi )+\pi ^2 \chi +3 (\chi +1) \log (\chi ) \log (\chi +1) \,. \la{fEq}
\eeq 
\item $p_2(z)$ is a simple constant:
\begin{equation}
\begin{split}
p_2 &= \frac{\mathbb{T}}{24\pi^2}\bigg(6 (2 \chi +1) \text{Li}_2(-\chi )+2 \pi ^2 \chi -6 (\chi +1) \log (\chi +1)\\
&\hspace{2cm}-3 \log \left(\frac{1}{\chi
   }\right) (2 \chi +(2 \chi +1) \log (\chi +1))\bigg)
\end{split}
\end{equation}
\item $\alpha_2(z,\bar z)$ is a complicated function built out of dilogarithms and simpler. In the real line, however, $\alpha$ is still given by Liouville which means that $\alpha_2$ in the line is again a simple constant!
\beq
 \alpha_2(z,z)=2\,p_2 \,.  
\eeq
\end{itemize}
What about the area? Can we integrate $\exp\!\big({\alpha_0+\tfrac{1}{L^2} \alpha_2}\big)$ over the upper half plane to obtain the first correction to the GM area? We can but it is painful. Here is the result for the full area of the minimal surface:
\begin{equation}
    \begin{split}
        \texttt{Area}_{AdS} &=\frac{1}{2\pi}\left[ \mathbb{S}\log\left(\frac{\mathbb{S}+\mathbb{T}}{\mathbb{S}}\right) + \mathbb{T}\log\left(\frac{\mathbb{S}+\mathbb{T}}{\mathbb{T}}\right)\right] + \frac{1}{L^2}\delta\texttt{Area},
    \end{split}
\end{equation}
where the leading term is recognized as the Gross-Manes area \cite{Gross:1987kza} and $\delta\texttt{Area}$ is the first correction to the GM area
\begin{equation}\label{eq:ads-area-correction}
    \begin{split}
        \delta \texttt{Area} &= \frac{1}{{48 \pi ^3}} \bigg[-6\text{Li}_2\left(-\frac{\mathbb{T}}{\mathbb{S}}\right) \left(\mathbb{S}^2 \log \left(\frac{\mathbb{S}+\mathbb{T}}{\mathbb{S}}\right)+\mathbb{T}^2 \log \left(\frac{\mathbb{T}}{\mathbb{S}+\mathbb{T}}\right)+2
   \mathbb{S} \mathbb{T} \log \left(\frac{\mathbb{T}}{\mathbb{S}}\right)\right)\\
   &\hspace{1.5cm}+18 \left(\mathbb{S} (\mathbb{S}+2 \mathbb{T}) \text{Li}_3\left(-\frac{\mathbb{T}}{\mathbb{S}}\right)+(\mathbb{S}-\mathbb{T}) (\mathbb{S}+\mathbb{T})
   \text{Li}_3\left(\frac{\mathbb{T}}{\mathbb{S}+\mathbb{T}}\right)+\mathbb{T}^2 \zeta (3)\right)\\
   &\hspace{1.5cm}+3 (\mathbb{T}-\mathbb{S}) (\mathbb{S}+\mathbb{T}) \log \left(\frac{\mathbb{S}+\mathbb{T}}{\mathbb{T}}\right) \log
   ^2\left(\frac{\mathbb{S}+\mathbb{T}}{\mathbb{S}}\right)\\
   &\hspace{1.5cm}+2 \pi ^2 \mathbb{T} \left(\mathbb{T} \log \left(\frac{\mathbb{T}}{\mathbb{S}}\right)+(2 \mathbb{S}-\mathbb{T}) \log
   \left(\frac{\mathbb{S}+\mathbb{T}}{\mathbb{S}}\right)\right)\bigg]. 
    \end{split}
\end{equation}
Checks of (\ref{fEq}) to (\ref{eq:ads-area-correction}) can be found in the ancillary file \texttt{ForPaper.nb}. 

It is remarkable that this classical worldsheet can be shown to match exactly the recent AdS amplitude calculation in \cite{Alday:2025pmg}. The full matching is fully discussed in Appendix \ref{app:correction-match}. It is worth emphasizing that the methods with which we derived this are distinct from the ones in \cite{Alday:2025pmg} in an important regard. In that reference, the authors considered the string amplitude in flat space written as a disk integral. While this comes from a worldsheet calculation, to add AdS corrections the authors postulate the same form of a disk integral and bootstrap the integrand corrections. Here, instead, we arrive at the same result by a classical worldsheet calculation in AdS, which further strengthens the worldsheet interpretation of the calculation in \cite{Alday:2025pmg} for the full AdS amplitude.

Finally, it is useful to spell out how $\delta\texttt{Area}$ connects to the coefficients introduced before. First, consider the symmetric kinematics $\mathbb S=\mathbb T$. Equation \eqref{eq:ads-area-correction} then reduces to \begin{equation} \delta\texttt{Area}(\mathbb T,\mathbb T)=\gamma\mathbb T^2,\qquad\gamma=\frac{\pi^2\log(16)-45\zeta(3)}{96\pi^3}.\end{equation}which matches exactly the correction around flat space found from the Nambu-Goto numerics. The conformal-gauge calculation therefore reproduces the value obtained independently from the numerical Nambu-Goto analysis as expected, since we already argued that the polygonal frames are equivalent. The improvement produced by this correction is displayed in Figure \ref{fig:gross-manes-correction}.
We can similarly take the Regge limit $\mathbb S\gg\mathbb T$. The $\log\mathbb{S}$ contribution to \eqref{eq:ads-area-correction} is\begin{equation}\left.\delta\texttt{Area}(\mathbb S,\mathbb T)\right|_{\log}=\alpha\mathbb T^2\log{\mathbb S},\qquad\alpha=-\frac{15+2\pi^2}{48\pi^3}.\end{equation}Thus the correction to Gross-Manes in this regime derived directly in conformal gauge agrees with the prediction obtained from the Regge solution, as illustrated by the small $T$ comparison in the left panel of Figure \ref{fig:regge-area-expansions}.
These two limits show that the general correction \eqref{eq:ads-area-correction} simultaneously reproduces the independent symmetric and Regge checks.

\section{Conclusions}

Our main result is arguably the plot in figure~\ref{fig:area-numerics} for the
area of a minimal surface ending on a null polygon in AdS. This minimal surface
interpolates beautifully between small solutions -- which effectively see flat
space and thus match the previous analysis of Gross and Manes~\cite{Gross:1989ge} 
-- and very large solutions, whose huge
boundary polygon is effectively pushed to the AdS boundary and which thus match
nicely the minimal surfaces of Alday and Maldacena~\cite{Alday:2007hr}. Several
of these limiting cases -- as well as another limit, the Regge limit -- were
studied analytically in the main text; figure~\ref{fig:roadmap} summarizes the
various connections between these results.

We see all these results both as \textit{starting points} and as
\textit{targets}.

Mathematically, they are \textit{starting points}. Null polygons in $AdS_3$ are
a measure-zero subset of all interesting polygons in all interesting spaces.
Can we compute space-like polygons in Euclidean AdS, perhaps starting from
\cite{kokubu1997weierstrass,aiyama2000kenmotsu}? 
Can we take the limit of a large number of edges and
thus approach general smooth curves as limits of polygons, as in
\cite{Toledo:2014koa}?

They are also \textit{targets}. The approach followed here was, in a way, a
brute-force approach: given the problem of minimizing a surface, we found the
surface -- sometimes numerically, sometimes analytically -- and then evaluated
its area. It would be fascinating to instead use the integrability of the
string sigma model to obtain these areas directly, without ever finding the
explicit shape of the minimal surface. As reviewed above, this is often
possible in closely related examples.

Within string theory and holography, such a putative integrability description
can hint at important physics beyond the classical limit. For instance, the
integrability description of the area of null polygons ending \emph{on} the AdS
boundary unveiled an operator product expansion for null polygonal Wilson loops
valid beyond the strong coupling limit~\cite{Alday:2010ku}, which in turn led
to a full finite-coupling solution in terms of the so-called pentagon
decomposition~\cite{Basso:2013vsa}. Our polygons contain those polygons as a
particular limit. Is there a generalization of the pentagons that would compute
our areas at any coupling?

This begs the obvious question: \textit{what are we computing exactly, and how
would we test any putative quantum conjecture?} Null polygons ending at the
boundary of AdS correspond to null polygonal Wilson loops, which are dual to
gluon scattering amplitudes \cite{Alday:2007hr,Drummond:2007aua,
Brandhuber:2007yx,Drummond:2008aq}. What do null polygons ending in the AdS
\emph{bulk} correspond to? (One answer -- if we are willing to use T-duality --
is W-boson scattering amplitudes \cite{McGreevy:2007kt,Alday:2009zm}; but what is their $T$-dual
counterpart?)

Ultimately, one long-term goal of these explorations is to extend holography.
If we understood the dual of these generalized null polygons ending in the
bulk, we would be able to interpolate between standard Wilson loops, dual to
big polygons, and some sort of ``flat-space Wilson loops'', dual to tiny
polygons. Can this teach us about flat-space holography? And could such a
flat-space holography also rely on string sigma model integrability as a
powerful tool for the study of its (string) amplitudes?

\section*{Acknowledgments}
It is a pleasure to thank Nima Arkani-Hamed, Elisabetta Armanini, Connor Behan, Simon Caron-Huot, Carolina Figueiredo, Davide Gaiotto, Eivind J{\o}rstad,  Rob Myers, Maria Nocchi, and Jon Toledo 
for useful discussions. We would like to give special thanks to Pinaki Banerjee for collaboration, encouragement, and many insightful discussions during an earlier stage of this project. 
Research at the Perimeter Institute is supported by the Government of Canada through the Department of Innovation, Science and Industry Canada, and by the Province of Ontario through the Ministry of Colleges and Universities.
 LG would like to acknowledge
the support provided by FAPESP Foundation through the grant 2023/04415-2.
SP is supported by the Simons Collaboration on Celestial Holography (MPS-CH-00001550) and the Celestial Holography Initiative at Perimeter.
PV is supported in part
by Discovery Grants from the Natural Sciences and Engineering Research Council of Canada, and by the Simons Foundation through the ``Nonperturbative Bootstrap'' collaboration (488661).

\appendix

\section{Deriving the Regge Solution}\label{app:regge-derivation}

In this appendix we will derive an exact solution to the minimal surface problem that covers the Regge limit $S\gg T$. From the Regge Lagrangian
\begin{equation}
    \mathbb{L} = \frac{ \sqrt{4 f(\eta ) (f(\eta )+1)-f'(\eta )^2}}{2 (f(\eta )+1)^{3/2}},
\end{equation}
valid in the large $S$ limit, we find that there is a conserved quantity
\begin{equation}
    \begin{split}
    \mathcal{E} &\equiv f'(\eta)\dfrac{\partial\mathbb{L}}{\partial f'(\eta)} - \mathbb{L}\\
    &=-\frac{f(\eta )}{\sqrt{f(\eta )+1} \sqrt{4 f(\eta ) (f(\eta )+1)-f'(\eta )^2}}.
    \end{split}
\end{equation}
We can then solve $\mathcal{E}(\eta) = \mathcal{E}(0)$ for $f'(\eta)$ by further demanding that for $\eta = 0$ we have $f(0) = f_0$ and $f'(0) = 0$. This gives
\begin{equation}
    f'(\eta) = -\frac{2 \sqrt{f(\eta ) \left(-\left(f_0^2+1\right) f(\eta )+f_0 f(\eta
   )^2+f_0\right)}}{\sqrt{f_0 (f(\eta )+1)}}.
\end{equation}
This makes it possible to find the inverse $\eta(f)$ instead of $f(\eta)$. In fact, by the inverse function theorem $\eta'(f)f'(\eta)=1$ and therefore we can find
\begin{equation}
    \begin{split}
        \eta'(f)&=-\dfrac{\sqrt{f_0(f+1)}}{2 \sqrt{f \left(-\left(f_0^2+1\right) f+f_0 f^2+f_0\right)}},
    \end{split}
\end{equation}
and now $\eta(f)$ follows from integrating over $f$ from $f_0$ to $f$
\begin{equation}
    \eta(f)=\frac{\Pi \left(\frac{1}{f_0+1};\arcsin \left(\sqrt{f_0}\right)|\frac{1}{f_0}\right)-\Pi
   \left(\frac{1}{f_0+1};\arcsin\left(\sqrt{\frac{f
   (f_0+1)}{f+1}}\right)|\frac{1}{f_0}\right)}{\sqrt{f_0+1}} .
\end{equation}
The area now follows directly from integrating the Lagrangian as a function of $f$.

\section{Gross-Manes in Nambu-Goto Form} \label{GMNGAp}

The goal of this appendix is to derive equation~\eqref{GMSequalT} starting from the parameterization~\eqref{Qvar}. As we approach the flat limit of the null quadrilateral, the leading term in~\eqref{Xpol} becomes
\beqa
\left(\begin{array}{c}
\mathbb{X}_1\\
\mathbb{X}_2\\
\mathbb{X}_3\\
\mathbb{X}_4 \end{array}\right)&=&\left(
\begin{array}{cccc}
 L & 0 & \frac{\sqrt{\mathbb{S}}}{2} & 0 \\
 L & \frac{1}{2} \sqrt{\mathbb{S}+\mathbb{T}} & 0 & \frac{\sqrt{\mathbb{T}}}{2} \\
L& 0 & -\frac{\sqrt{\mathbb{S}}}{2} & 0 \\
 L & \frac{1}{2} \sqrt{\mathbb{S}+\mathbb{T}} & 0 & -\frac{\sqrt{\mathbb{T}}}{2} \\
\end{array}
\right) \nn
\eeqa
so that
\beqa
\left(\begin{array}{c}
k_1\\
k_2\\
k_3\\
k_4 \end{array}\right)&=&\left(
\begin{array}{cccc}
 0 & \frac{1}{2} \sqrt{\mathbb{S}+\mathbb{T}}  & - \frac{\sqrt{\mathbb{S}}}{2} &\frac{\sqrt{\mathbb{T}}}{2} \\
 0 & -\frac{1}{2} \sqrt{\mathbb{S}+\mathbb{T}} & -\frac{\sqrt{\mathbb{S}}}{2} & -\frac{\sqrt{\mathbb{T}}}{2} \\
0 & \frac{1}{2} \sqrt{\mathbb{S}+\mathbb{T}}  & \frac{\sqrt{\mathbb{S}}}{2} & -\frac{\sqrt{\mathbb{T}}}{2} \\
 0 & -\frac{1}{2} \sqrt{\mathbb{S}+\mathbb{T}} & \frac{\sqrt{\mathbb{S}}}{2} & \frac{\sqrt{\mathbb{T}}}{2} \\
\end{array}
\right) \nn
\eeqa
where $k_i=(\mathbb{X}_{i+1}-\mathbb{X}_i)$ which are now null vectors in an $\mathbb{R}^{1,2}$ subspace of the embedding $\mathbb{R}^{2,2}$. Now we would like to find a set of dual vectors $q_i$ such that
\be
q_i\cdot k_j=\delta_{ij},~~i,j\in\{1,2,3\}.
\ee
Note that $\{k_1,k_2,k_3\}$ span $\mathbb{R}^{1,2}$ while  $k_1+k_2+k_3+k_4$=0, as expected since the null polygon is closed. This implies that finding such a $q$ just amounts to inverting the above equation as a matrix
\be\label{eq:orth}
q_{i\mu}k_{~j}^{\mu}=\delta_{ij}
\ee
for $\mu$ restricted to the last three components. Namely 
\beqa
\left(\begin{array}{c}
\vec{q}_1\\
\vec{q}_2\\
\vec{q}_3\\
 \end{array}\right)&=&\left(
\begin{array}{cccc}
-\frac{1}{ \sqrt{\mathbb{S}+\mathbb{T}}}  & -\frac{1}{\sqrt{\mathbb{S}}} & 0 \\
0& -\frac{1}{\sqrt{\mathbb{S}}} & -\frac{1}{\sqrt{\mathbb{T}}}  \\
-\frac{1}{ \sqrt{\mathbb{S}+\mathbb{T}}}  & 0 &  -\frac{1}{\sqrt{\mathbb{T}}} \\
\end{array}
\right). \nn
\eeqa
Now the $q_{i,0}$ component is unconstrained thus far.
Looking back at~\eqref{Xpol} recall that we have
\be
X_\texttt{polygon}(z,\bar z)=\mathbb{X}_1+\sum_{i=1}^4 (\mathbb{X}_{i+1}-\mathbb{X}_i) \mathcal{L}_i \,, \qquad \mathcal{L}_j \equiv \frac{1}{2\pi i} \Big(\log\Big( -\frac{z-\sigma_j}{\bar z-\sigma_j}\Big) - i \pi \Big) \ee
We can now define 
\beq
Q_i\equiv \exp(2\pi i \,q_i \cdot X_\texttt{polygon})=e^{2\pi i q_i\cdot \mathbb{X}_1}\frac{z-\sigma_i}{\bar z-\sigma_i}\Big/ \frac{z-\sigma_{4}}{\bar z-\sigma_{4}},\quad 1\leq i\leq 3
\label{Qvar2}
\eeq 
where momentum conservation gives the $\sigma_4$ dependence. As done in the main text, $Q_4 = 1$.
We see that we can pick $-q_{i0}L=\vec{q}_i\cdot \vec{\mathbb{X}}_1$ to kill the overall phase. 

At the same time we can go back to the embedding space parameterization we were using for the Nambu-Goto story in section~\ref{sec:nambu-goto-numerics}. We must be careful that, as already discussed in section \ref{sec:flat-space-corrections} the frames used here and there are distinct and related by the transformation $\Lambda$ defined in \eqref{eq:rotation-frame-sec-3-to-4}. Letting $P$ be given by \eqref{eq:Xemb} in the frame of section \ref{sec:nambu-goto-numerics} we have that $\Lambda P$ becomes in the flat space limit 
\be\badat{3}\label{eq:flat-limit-Xemb}
X^\mu&=(L, \tau_0,\frac{1}{4}\sqrt{\mathbb{S}}(x+y),\frac{1}{4}\sqrt{\mathbb{T}}(y-x))) .
\eadat\ee
where we've defined $\tau_0=\frac{\sqrt{\mathbb{S}+\mathbb{T}}}{4}-\tau L$. We see that $q_i\cdot X$ gives a linear system we can invert to solve for
\beqa
\left(\begin{array}{c}
\tau_0\\
x\\
y\\
 \end{array}\right)&=&\left(\begin{array}{c} \frac{\sqrt{\mathbb{S}+\mathbb{T}}}{2}(q_1\cdot X-q_2\cdot X+q_3\cdot X) \\
  -2q_1\cdot X+2q_3\cdot X+1  \\
  -2q_2\cdot X+1 
 \end{array}\right)
\eeqa
where the shifts come from canceling the phase in~\eqref{Qvar2}. 
We can now write
\be
e^{\frac{4\pi i \tau_0}{\sqrt{\mathbb{S}+\mathbb{T}}}}=\frac{Q_1 Q_3}{Q_2},~~e^{i\pi x}=-\frac{Q_3}{Q_1},~~e^{i\pi y}=-\frac{1}{Q_2}.
\ee
For fixed puncture locations this surface is just parameterized by $(z,\bar z)$ and we can use this to solve for $\tau_0(x,y)$. 
As discussed above
 \beq
\frac{(Q_a-Q_b)(Q_c-Q_d)}{(Q_a-Q_c)(Q_b-Q_d)}=\frac{(\sigma_a-\sigma_b)(\sigma_c-\sigma_d)}{(\sigma_a-\sigma_c)(\sigma_b-\sigma_d)}=\texttt{constant}_{abcd} 
\eeq
while the Virasoro constraint gives \be
\frac{(\sigma_1-\sigma_2)(\sigma_3-\sigma_4)}{(\sigma_2-\sigma_3)(\sigma_1-\sigma_4)}=\frac{\mathbb{S}}{\mathbb{T}}.
\ee
Noting that
\be
2\cos\frac{4\pi\tau_0}{\sqrt{\mathbb{S}+\mathbb{T}}}=\frac{Q_1Q_3}{Q_2}+\frac{Q_2}{Q_1Q_3},~~2\cos\pi x=-\frac{Q_3}{Q_1}-\frac{Q_1}{Q_3},~~2\cos\pi y=-\frac{1}{Q_2}-{Q_2}
\ee
Using the relation above with $\mathbb{Q}_4=1$ and solving for $\mathbb{Q}_3$ in terms of the cross ratio, we see that when $\mathbb{S}=\mathbb{T}$ we indeed have
\be
\frac{1}{2}(\cos (\pi  x) (\cos (\pi  y)-1)-\cos (\pi
    y)-1)=\cos\left(\frac{\sqrt{2}}{R}\pi \tau_{0}\right) 
\ee
where $R=\sqrt{\mathbb{S}}/2$. Finally using that $\tau_0 = \frac{\sqrt{2\mathbb{S}}}{4}-\tau L$, we find
\be
\frac{1}{2}(\cos (\pi  x) (\cos (\pi  y)-1)-\cos (\pi
    y)-1)=-\cos\left(\frac{\sqrt{2}}{R}\pi \tau L\right). 
\ee
This equation determines only the cosine of the time coordinate and therefore leaves a branch ambiguity. On the principal branch, and imposing continuity together with the boundary condition, the appropriate choice in each quadrant is
\begin{equation}
    \tau(x,y) = \frac{\sqrt{\mathbb{S}}}{\sqrt{2}\pi L}\arcsin\left[\sin\left(\frac{\pi x}{2}\right)\sin\left(\frac{\pi y}{2}\right)\right]
\end{equation}
and once we recall that $\mathbb{S}=L^2S$ for the $S$ in section \ref{sec:nambu-goto-numerics} we find perfect agreement with ~\eqref{GMSequalT}. 

Following the same logic for $S\neq T$ would lead to the general Gross-Mende solution in NG form as 
\beq
\tau(x,y) = \dfrac{\sqrt{S+T}}{2\pi L}\arcsin \left(\frac{T\cos \tfrac{\pi}{2}(x-y)-S\cos\tfrac{\pi}{2}(x+y)}{S+T}\right)
\eeq

\section{RBI Calculation of the Gross-Manes Area} \label{RBIAp} 

In this appendix we employ the Riemann Bilinear identity (RBI) to compute the area of the Gross-Manes minimal surface. In general we know that for the $n$-point minimal surface spanning a null polygon,
\begin{equation}
    p(z) = \dfrac{P_{n-4}(z)}{\prod_{i=1}^n (z-\sigma_i)},
\end{equation}
where $P_{n-4}(z)$ is a degree $n-4$ polynomial and where $\{\sigma_i\}$ are subject to the scattering equations. We are going to parameterize the momenta by
\begin{equation}
p_i = \omega_i (1+z_i^2,-1+z_i^2,2z_i),
\end{equation}
where the metric is $\eta = \operatorname{diag}(-1,1,1)$. The simplest solution to the scattering equations is given by $\sigma_i = z_i$ and in that case $p(z)$ admits a very simple form that can be derived directly from the Gross-Manes embedding with this choice of $\sigma_i$:
\begin{equation}
    p(z) = \frac{1}{\pi i}\sum_i \dfrac{\omega_i}{z-z_i}.
\end{equation}
In fact, for the Gross-Manes solution with $n$ punctures at positions $\sigma_i$ and solution to the scattering equations $z_i = \sigma_i$, the Weierstrass data $\mu(z)$ and $\nu(z)$ of the embedding is given by\footnote{The Weierstrass data is defined by solving the Virasoro condition $\partial X\cdot\partial X=0$ in terms of two unconstrained functions $\mu(z)$ and $\nu(z)$ as $$\partial X = \frac{\mu}{2}(1+\nu^2,-1+\nu^2,2\nu).$$}:
\begin{align}
    \nu(z) &= z, \\
    \mu(z) & = \frac{1}{\pi i}\sum_i\frac{\omega_i}{z-z_i}.
\end{align}
Moreover, $p(z)$ can be written in terms of the Weierstrass data as $p(z) = \mu(z)\nu'(z)$, which in this case reduces to
\begin{equation}
    p(z) = \frac{1}{\pi i}\sum_i\frac{\omega_i}{z-z_i}
\end{equation}
We will henceforth assume this solution to the scattering equations and this form of $p(z)$. The extension to the more general case is straightforward. Let $\Sigma\subset \mathbb{R}^{1,2}$ be the minimal surface described by the Gross-Manes embedding. The area of the surface is given by the functional:
\begin{equation}
    \texttt{Area}(\Sigma) = -\dfrac{1}{2i}\left(\int_{\Sigma} (\nu(z) - \overline{\nu}(\overline{z}))^2 \mu(z) dz \wedge \overline{\mu}(\overline{z}) d\overline{z}\right).
\end{equation}
Substituting the Weierstrass data into the area functional and expanding the quadratic term $(z - \overline{z})^2$, we obtain a decomposition into factorized integrands:
\begin{equation}\label{eq:area-from-forms-rbi}
    \texttt{Area}(\Sigma) = -\dfrac{1}{2i}\left( \int_{\Sigma} \omega_2 \wedge \overline{\omega}_0 - 2 \int_{\Sigma} \omega_1 \wedge \overline{\omega}_1 +\int_{\Sigma} \omega_0 \wedge \overline{\omega}_2\right),
\end{equation}
where we have defined the family of one-forms\footnote{These are coordinate-dependent definitions because $p(z)$ is a component of a quadratic differential instead of a one-form, and hence does not transform properly. The actual invariant one-forms should be thought of as having component in any other overlapping coordinate system given by the appropriate transformation of the component defined in this system. It would be interesting to perform a RBI calculation of the area that uses the actual one-form $\sqrt{p}dz \in \Omega_{\tilde{\Sigma}}^1$.}
\begin{equation}
    \omega_m \equiv z^m p(z) dz, \quad m \in \{0, 1, 2\},
\end{equation}
which as we will see momentarily, are really one-forms in a double cover $\tilde{\Sigma}$ after regularization. Moreover, the embedding is initially defined on the upper-half plane so that $\Sigma\simeq \mathbb{H}$. Before constructing the double cover we double the surface to the sphere passing to $\hat{\Sigma}\simeq \mathbb{CP}^1$, the so-called Schottky double. This merely introduces a factor of $2$ in the area:
\begin{equation}
    \texttt{Area}(\Sigma) = \frac{1}{2}\texttt{Area}(\hat{\Sigma})
\end{equation}
Finally, to invoke the Riemann Bilinear Identity the surface over which we integrate must have non-trivial genus. Here the forms have simple poles and are defined on a genus $g = 0$ domain $\hat{\Sigma}$. We follow the regularization procedure used before e.g. in \cite{Caetano:2012ac} that amounts to blowing up the simple poles into branch cuts. Specifically,
\begin{equation}
    \dfrac{1}{z-\sigma_i}\mapsto \dfrac{1}{\sqrt{(z-\sigma_i)(z-\sigma_i-\epsilon)}},\quad \epsilon\to 0
\end{equation}

After the regularization the forms are defined on a particular hyperelliptic double covering curve $\tilde{\Sigma}$ of $\hat{\Sigma}$. By the Riemann-Hurwitz theorem, this is a genus $g = n-1$ surface, and the area integral lifts with a factor of $1/2$ since $\tilde{\Sigma}$ is a double cover of $\hat{\Sigma}$. Altogether we have
\begin{equation}
    \texttt{Area}(\Sigma) = \frac{1}{4} \int_{\tilde{\Sigma}} \pi^* \Omega,
\end{equation}
where $\Omega$ is the form defined in \eqref{eq:area-from-forms-rbi} pulled back to the double cover $\tilde{\Sigma}$ and the factor of $4$ accounts both from doubling to the sphere and lifting to the double cover.

We are now ready to apply the Riemann Bilinear Identity. For any two closed 1-forms $\alpha$ and $\beta$ on $\tilde{\Sigma}$ with a symplectic homology basis $\{A_k, B_k\}_{k=1}^g$, we have:
\begin{equation}
    \int_{\tilde{\Sigma}} \alpha \wedge \overline{\beta} = \sum_{k=1}^g \left( \oint_{A_k} \alpha \oint_{B_k} \overline{\beta} - \oint_{B_k} \alpha \oint_{A_k} \overline{\beta} \right).
\end{equation}
Applying this to our area decomposition:
\begin{equation}\label{eq:area-from-rbi-1}
    \texttt{Area}(\Sigma) = -\frac{1}{8i} \sum_{k=1}^{n-1} \sum_{i=0}^2 (-1)^i\binom{2}{i} \mathcal{Q}_k(\omega_{2-i},\omega_i),
\end{equation}
where the bilinear pairing $\mathcal{Q}_k$ is defined as:
\begin{equation}
    \mathcal{Q}_k(\omega_a, \omega_b) \equiv \oint_{A_k} \omega_a \overline{\oint_{B_k} \omega_b} - \oint_{B_k} \omega_a \overline{\oint_{A_k} \omega_b}.
\end{equation}
\begin{figure}\label{ab-cycles}
        \centering
        \includegraphics[width=0.5\linewidth]{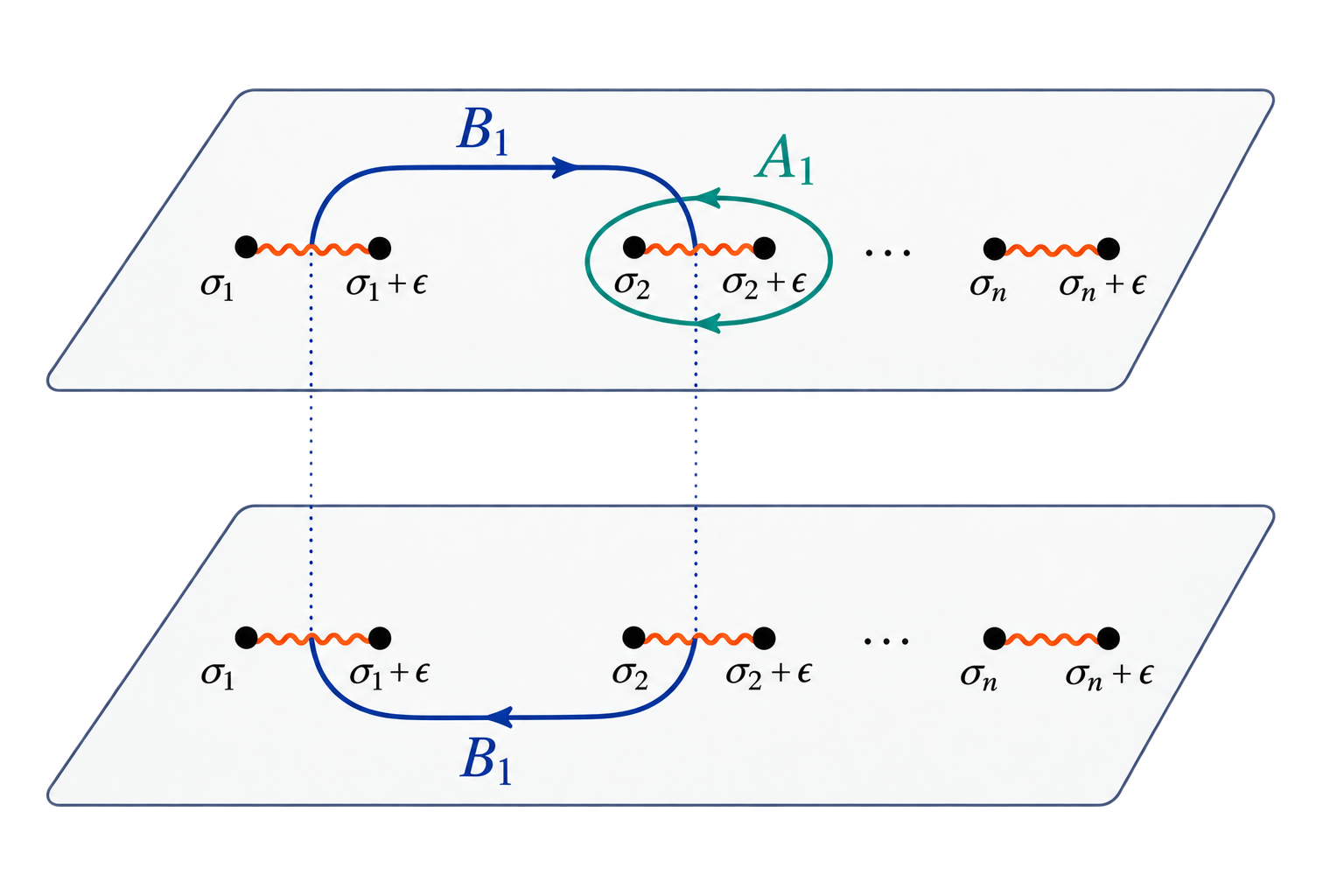}
        \caption{The definition of $A$ and $B$ cycles used to compute the area through RBI after regularization. The $A_k$ cycle encircles the infinitesimal cut at $\sigma_{k+1}$, whereas the $B_k$ cycle starts from the cut at $\sigma_1$ goes to the second sheet through the cut at $\sigma_{k+1}$ and then comes back through the cut at $\sigma_1$ to the first sheet.
        } 
    \end{figure}
We denote the residue of the form $\omega_m$ at the puncture $\sigma_j$ as $r_{j,m}$. Letting $R_j \equiv \text{Res}_{z=\sigma_j} p(z)$, we have:
\begin{equation}
    r_{j,m} = \text{Res}_{z=\sigma_j} (z^m p(z)) = \sigma_j^m R_j.
\end{equation}
We choose the cycles as follows: $A_k$ encircles the cut at $\sigma_{k+1}$, and $B_k$ runs from the cut at $\sigma_1$ to $\sigma_{k+1}$ on the first sheet and returns on the second. The definition of the cycles is shown in Figure \ref{ab-cycles}. The A-cycle integral is determined by the residue at the enclosed puncture $\sigma_{k+1}$:
\begin{equation}
    \mathcal{A}_{k,m} \equiv \oint_{A_k} \omega_m = 2\pi i \, r_{k+1, m} = 2\pi i \, \sigma_{k+1}^m R_{k+1}.
\end{equation}
The B-cycle integral traverses the real axis between cuts. Since $\omega_m$ changes sign on the second sheet, the contributions add constructively:
\begin{equation}
    \mathcal{B}_{k,m} \equiv \oint_{B_k} \omega_m = 2 \int_{\sigma_1}^{\sigma_{k+1}} z^m p(z) dz.
\end{equation}
Using the partial fraction decomposition $z^m p(z) = \sum_{j=1}^n \frac{\sigma_j^m R_j}{z-\sigma_j}$, we integrate:
\begin{equation}
    \mathcal{B}_{k,m} = 2 \sum_{j=1}^n \sigma_j^m R_j \log \left( \frac{\sigma_{k+1} - \sigma_j}{\sigma_1 - \sigma_j} \right).
\end{equation}
Henceforth we shall denote $\Delta_{kj} \equiv \frac{\sigma_{k+1} - \sigma_j}{\sigma_1 - \sigma_j}$ to simplify the notation. We thus have
\begin{equation}
    \mathcal{B}_{k,m} = 2 \sum_{j=1}^n \sigma_j^m R_j \log \Delta_{kj}.
\end{equation}
The necessary bilinear pairings can then be written as
\begin{equation}
    \begin{split}
    {\cal Q}_k(\omega_{2-i},\omega_i) &= {\cal A}_{k,2-i} \overline{{\cal B}_{k,i}} - {\cal B}_{k,2-i} \overline{{\cal A}_{k,i}}\\
    &= \left(2\pi i\sigma_{k+1}^{2-i} R_{k+1} \right)\left(2 \sum_{j=1}^n \sigma_j^i \overline{R}_j \overline{\log \Delta_{kj}} \right) -\left(2 \sum_{j=1}^n \sigma_j^{2-i} R_j \log \Delta_{kj}\right) \left(-2\pi i \sigma_{k+1}^{i}\overline{R}_{k+1}\right)\\
    &=4\pi i \sum_{j=1}^n \left(\sigma_{k+1}^{2-i}\sigma_j^{i} R_{k+1}\overline{R}_j \overline{\log \Delta_{kj}}+\sigma_{k+1}^i \sigma_j^{2-i} R_j\overline{R}_{k+1} \log \Delta_{kj}\right)\\
    &=-4\pi i R_{k+1}\sum_{j=1}^n R_j\left(\sigma_{k+1}^{2-i}\sigma_j^{i} \overline{\log \Delta_{kj}}+\sigma_{k+1}^i \sigma_j^{2-i}\log \Delta_{kj}\right),
    \end{split}
\end{equation}
where the last line follows from the fact that the residues are purely imaginary. It then turns out that once we sum over $i$ weighted by the binomial coefficients and signs in \eqref{eq:area-from-rbi-1}, there is a great simplification:
\begin{equation}
    \sum_{i=0}^2 (-1)^i \binom{2}{i} \mathcal{Q}_k(\omega_{2-i}, \omega_i) = -4\pi i R_{k+1} \sum_{j=1}^n R_j (\sigma_{k+1} - \sigma_j)^2 \left(\log \Delta_{kj} + \overline{\log \Delta_{kj}}\right).
\end{equation}
Finally, recalling that the residues are related to the energies by $R_i = \frac{\omega_i}{\pi i}$, we obtain
\begin{equation}
    \sum_{i=0}^2 (-1)^i \binom{2}{i} \mathcal{Q}_k(\omega_{2-i}, \omega_i) = \frac{8i}{\pi} \sum_{j=1}^n \omega_j\omega_{k+1} (\sigma_{k+1} - \sigma_j)^2 \operatorname{Re}\log \Delta_{kj},
\end{equation}
which using the parameterization $p_i = \omega_i (1+\sigma_i^2,-1+\sigma_i^2,2\sigma_i)$ and metric $\eta = \operatorname{diag}(-1,1,1)$, and explicitly writing $\Delta_{kj}$, becomes:
\begin{equation}
    \sum_{i=0}^2 (-1)^i \binom{2}{i} \mathcal{Q}_k(\omega_{2-i}, \omega_i) = -\frac{4i}{\pi} \sum_{j=1}^n p_j\cdot p_{k+1} \left(\log|\sigma_j-\sigma_{k+1}| -\log |\sigma_1 - \sigma_{j}| \right)
\end{equation}
Finally the sum over $k$ together with momentum conservation -- or equivalently, the closure of the polygonal contour -- removes the dependence on the arbitrary reference puncture $\sigma_1$ and yields the Gross-Manes area
\begin{equation}
    \texttt{Area}(\Sigma) =\frac{1}{2\pi} \sum_{k\neq j}p_j\cdot p_{k} \log|\sigma_j-\sigma_{k+1}|.
\end{equation}

\section{Comparing Flat Space Corrections}\label{app:correction-match}

In this appendix, we are going to compare our results for the leading AdS corrections to the Gross-Manes area with the recent results that appeared in \cite{Alday:2025pmg}, derived by different methods. To that end, we must first and foremost clarify what the relation is between what we have computed and what they did in order to clarify exactly where a match should be expected.

To properly establish the connection, the worldsheet approach of \cite{Alday:2025pmg} starts from the flat space tree-level amplitude written as a disk integral with a correction to the integrand \cite{koba1969reaction}
\begin{equation}\label{eq:veneziano-flat-space}
    {\cal M}_{\rm AdS} = \frac{1}{S+T}\int_0^1 \dfrac{dx}{x(1-x)} x^{-S}(1-x)^{-T}\left(1+\frac{S+T}{L^2}h(S,T;x)\right)+O(L^{-4})
\end{equation}
from which $h(S,T;x)$ is fixed by bootstrap techniques. To compare to a classical world-sheet area we must study the high-energy limit $S\to\infty$ with $\frac{T}{S}$ fixed. In this scenario we can compute the amplitude via a saddle-point approximation by rewriting $x^{-S}(1-x)^{-T} = e^{-S_{\rm eff}}$ and extremizing this effective action $S_{\rm eff}(x) = S\log x+T\log(1-x)$. The saddle point is found to be at
\begin{equation}
    x^\ast = \dfrac{S}{S+T}.
\end{equation}
We note, in passing, that in principle the saddle gets a correction due to $h(S,T;x)$, but this saddle correction does not affect the area we are going to extract from this amplitude analysis at $O(L^{-2})$. Evaluating $S_{\rm eff}(x^\ast)$ we recover the area of the Gross-Manes minimal surface to leading order. More fundamentally, in this classical high-energy limit, the full AdS amplitude is the exponential of the on-shell Polyakov action with appropriate vertex operators evaluated  at the classical saddle, which by means of  a $T$-duality argument \cite{McGreevy:2007kt} becomes the area of a minimal surface with a null polygonal contour in AdS:
\begin{equation}{\cal M}_{\rm AdS} \simeq e^{-\texttt{Area}_{\rm AdS}}\end{equation}
and therefore expanding the area as $\texttt{Area}_{\rm AdS}= \texttt{Area} + \frac{1}{L^2}\delta \texttt{Area}$ we find
\begin{equation}{\cal M}_{\rm AdS} \simeq e^{-\texttt{Area}}\left(1-\frac{1}{L^2}\delta\texttt{Area}\right).\end{equation}
Comparison between this and the disk integral suggests that we should observe a match for the area correction we found and the function $h(S,T;x)$ bootstrapped by \cite{Alday:2025pmg} evaluated at the classical saddle $x^\ast$:
\begin{equation}\delta\texttt{Area} = -N h\left(S,T;\frac{S}{S+T}\right)\end{equation}
where $N$ is a possible normalization factor accounting for differing conventions. We can remarkably check that this matching can be observed exactly in the high-energy limit! The direct calculation of the minimal area correction gave the following result
\begin{equation}
    \begin{split}
        \delta\texttt{Area} &= \frac{1}{{48 \pi ^3}} \bigg[-6\text{Li}_2\left(-\frac{T}{S}\right) \left(S^2 \log \left(\frac{S+T}{S}\right)+T^2 \log \left(\frac{T}{S+T}\right)+2
   S T \log \left(\frac{T}{S}\right)\right)\\
   &\hspace{1.5cm}+18 \left(S (S+2 T) \text{Li}_3\left(-\frac{T}{S}\right)+(S-T) (S+T)
   \text{Li}_3\left(\frac{T}{S+T}\right)+T^2 \zeta (3)\right)\\
   &\hspace{1.5cm}+3 (T-S) (S+T) \log \left(\frac{S+T}{T}\right) \log
   ^2\left(\frac{S+T}{S}\right)\\
   &\hspace{1.5cm}+2 \pi ^2 T \left(T \log \left(\frac{T}{S}\right)+(2 S-T) \log
   \left(\frac{S+T}{S}\right)\right)\bigg]
    \end{split}
\end{equation}
and we found a precise match with the leading behavior of $h(S,T;\tfrac{S}{S+T})$ in the $S\to\infty$ limit with $\frac{T}{S}$ fixed and with $N = \frac{1}{2\pi^2}$. The detailed match is shown in the Mathematica notebook $\texttt{AreaCorrectionComparison.nb}$ that accompanies the arXiv submission.

 \bibliographystyle{utphys}
 \bibliography{references}

\end{document}